\documentclass[12pt]{iopart}
\usepackage{graphicx}
\usepackage{epstopdf}
\usepackage{natbib}

\newcommand{\be}{\begin{equation}}
\newcommand{\ee}{\end{equation}}
\newcommand{\bea}{\begin{eqnarray}}
\newcommand{\eea}{\end{begin}}
\newcommand{\p}{\partial}
\def\eg{{\rm e.g.$\,$}}

\input epsf

\begin{document}

\title{Observational signatures of Jordan-Brans-Dicke theories of gravity}

\author[Acquaviva \& Verde]{Viviana Acquaviva$^{1,2}$,Licia Verde$^{1,2}$}
\address{$^1$ Dept. of Physics and Astronomy, University of Pennsylvania, 209 South 33rd Street, Philadelphia, PA 19104}
\address{ $^2$ Dept. of Astrophysical Sciences, Peyton Hall, Princeton University, Princeton, NJ 08540}
\ead{vacquavi@princeton.edu,lverde@physics.upenn.edu}

\begin{abstract} 
We analyze the Jordan-Brans-Dicke model (JBD) of gravity, where deviations from General Relativity (GR) are described by a scalar field non-minimally coupled to the graviton. The theory is characterized by a constant coupling parameter, $\omega_{\rm JBD}$; GR is recovered in the limit $\omega_{\rm JBD} \rightarrow \infty$. In such theories, gravity modifications manifest at early times, so that one cannot rely on the usual approach of looking for inconsistencies in the expansion history and perturbations growth in order to discriminate between JBD and GR. However, we show that a similar technique can be successfully applied to early and late times observables instead. Cosmological parameters inferred extrapolating early-time observations to the present will match those recovered from direct late-time observations only if the correct gravity theory is used. We use the primary CMB, as will be seen by the Planck satellite, as the early-time observable; and forthcoming and planned Supernov{\ae}, Baryonic Acoustic Oscillations and Weak Lensing experiments as late-time observables. We find that detection of values of $\omega_{\rm JBD}$ as large as 500 and 1000 is within reach of the upcoming (2010) and next-generation (2020) experiments, respectively.
\end{abstract}

\section{Introduction}
Theoretical modifications to General Relativity (GR) were in many cases formulated much before
the discovery of cosmic acceleration, often within the more general
quest of the search for a parent theory of GR which
could lead to the Grand Unification of the four fundamental forces (e.g. \cite{gsw,adelberger} and references therein). 
In the last few years, however, possible applications of modified
gravity theories as an explanation for cosmic acceleration have
generated a renewed interest in the cosmological community towards
such models, and many new ones have been formulated  (\eg \cite{Carroll03,Capozziello:2002rd,Capozziello:2003tk,Nojiri:2003ft,Nojiri:2003ni,Dolgov:2003px,Hu:2007nk,DGP,DDG02} and references therein).
The most compelling shortcoming of the standard Cold Dark Matter model is the
observed mismatch in the right and left hand side of Einstein
equations, which can be mostly simply accounted for through a
Cosmological Constant term. However, the form of such equations is
derived from the Lagrangian of GR, so that an
alternative explanation is that the latter breaks down on cosmological scales.
In this paper we analyze possible observational signatures of a class
of modified GR models known as scalar-tensor theories (\eg \cite{Damour:1992we,Uzan:2002vq,Fuji_Maeda}), where the
action of gravity is determined by a scalar field in addition to the metric tensor. 
We consider the simplest example of scalar-tensor theory, the Jordan-Brans-Dicke model (JBD), \cite{Brans:1961sx,Peebles:1970ag}; in this case, the
coupling between gravity and the scalar field is described by a constant parameter, $\omega_{\rm JBD}$.
The main motivation for this work is to provide a general approach to test
GR on cosmological scales, as opposed to Solar System
scale measurements; this is in principle a well distinct task from
that of addressing the cosmic acceleration problem. 
In particular, as we will show later, in the JBD model
the modifications of gravity are limited to early times, so that acceleration
cannot be obtained as a result of the gravity modification alone. 
However, the theory predicts the introduction of a scalar field as a gravitational degree of freedom,
and such scalar field behaves as minimally coupled at late times, so that it can play the role of Quintessence, given a suitable potential. In this sense, the JBD model can provide an explanation of the nature of Dark Energy \cite{Uzan:2006mf}; it does not address the fine tuning problem, which is, however, common to most dark energy models. \\
The strongest constraint to date on this model has been put on the Solar System scale: the present $2\sigma$ limit from the Cassini spacecraft is $\omega_{\rm JBD} > 40000$ \cite{Bertotti:2003rm}. However, such constraint does not necessarily apply on distances much larger than those of the measurements, and epochs much different from the present. Local universe experiments only probe scales in gravitational equilibrium, where the background expansion of the Universe is negligible; they would not reveal spatial or time variation of the gravitational constant on larger scales \cite{Barrow:1999qk,Clifton:2004st}. Sensible limits on the value of $\omega_{\rm JBD}$ which is representative of the whole Universe have to be inferred from observations on cosmologically relevant scales \cite{Shaw:2005vf}.
The current limit on such quantity, obtained combining the extended WMAP 1st year data and the 2dF large-scale structure data, is $\omega_{\rm JBD} > 120$ at 95$\%$ confidence level \cite{Acquaviva:2004ti}. \\
In the present paper we will show how constraints on the ``cosmological'' JBD parameter can be substantially improved using suitable combinations of next-generation experiments. We consider the CMB power spectra, coupled at a time with Supernov{\ae} type Ia (SNe), Baryon Acoustic Oscillations (BAO), and Weak Lensing (WL).  The main difference with respect to earlier works is that we will not assume previous knowledge of the model. We propose a general method for discriminating between GR and JBD, which relies on the fact that the two models are different at early times and similar at late times. Therefore, observations at early times, if extrapolated to the present epoch, will be consistent with late times observations only if the correct theory of gravity is used. This is somehow close in spirit to the consistency checks between expansion history and perturbations growth often used in order to discriminate between dark energy and modifications to GR,  in the case that they differ at late epochs (\eg \cite{Lue1,Linder:2005in,Koyama:2005kd,Knox_Song_2,Upadhye,Ishak_Spergel,Huterer:2006mv,Kunz:2006ca,Bludman:2007kg,Chiba:2007rb}, and references therein). \\
The method presented in this paper can be easily applied to any pair of models which agree at some epoch and disagree at some other epoch. This include not only modifications to GR as opposed to GR \cite{Amendola:2006kh,Agarwal:2007wn}, but also, for example, Ordinary \cite{Wetterich:1987fm,Ratra:1987rm,Caldwell:1997ii} versus Early \cite{Hebecker:2000zb,ArmendarizPicon:2000dh,Albrecht:1999rm,Wetterich:2003jt}, or Extended \cite{ext_quint,track_extquint,Matarrese:2004xa}, Quintessence models. \\
The outline of the paper is as follows. In Sec. \ref{outline} we present the JBD model and its general phenomenology. In Sec. \ref{early} we compare a dark-fluid GR model with the same expansion history and the JBD model, showing how the JBD field behaves in terms of equation of state and how it would change the interpretation of the cosmological parameters if Cosmology was assumed to be GR. In Sec. \ref{earlylate} we present our method in detail, describe the observables that we are going to use, and present results for two values of $\omega_{JBD}$ of interest. Finally, in Sec. \ref{conclu} we summarize our results and discuss their implications.

\section{Phenomenology of the JBD model} 
\label{outline}
 The JBD cosmological model was formulated in 1961 as as the first scalar-tensor theory of gravity \cite{Brans:1961sx}. It only features one more degree of freedom with respect to GR, the JBD parameter $\omega_{\rm JBD}$, which is constant both in space and time. The Lagrangian of the JBD model reads
\be
{\cal L_{\rm JBD}} = \frac{1}{16\pi G}\left({\bf \Phi}R - \frac{\omega_{\rm JBD}}{\bf \Phi} \p_\mu {\bf \Phi}\p^\mu {\bf \Phi}\right) - V({\bf \Phi}) + {\cal L}_{\rm fluid}
\ee
where R is the Ricci scalar, and ${\cal L}_{\rm fluid}$ is the Lagrangian of the ordinary matter and radiation components, . Its equation of motion are
\be
{H}^2 + {H} \frac{\dot{\bf \Phi}}{\bf \Phi} = \frac{\omega_{\rm JBD}}{6}\left(\frac{\dot{\bf \Phi}}{\bf \Phi}\right)^2 + \frac{8 \pi G}{3}\frac{\rho}{\bf \Phi}\,; 
\ee
\be
\ddot{\bf\Phi} + 3 {H} \dot{\bf \Phi} = \frac{8 \pi G}{2 \omega_{\rm JBD}+3}(\rho - 3 p)\;,
\ee
here $H$ is the Hubble parameter $\dot{a}/a$, where $a$ is the scale factor, and dots denote derivatives with respect to proper time.  \\
GR is a particular case of the JBD theory, corresponding to $\omega_{\rm JBD}  = \infty$. For such value the solution of the above equations is ${\bf \Phi}$ = cost = 1, and ${\bf \dot\Phi/\Phi} \rightarrow 0$ steeply, so that additional terms in the Friedmann equation disappears and its ordinary gravity form is recovered \cite{Will}.

\subsection{Background dynamics}
\begin{figure}
\centering
\includegraphics[width=5cm]{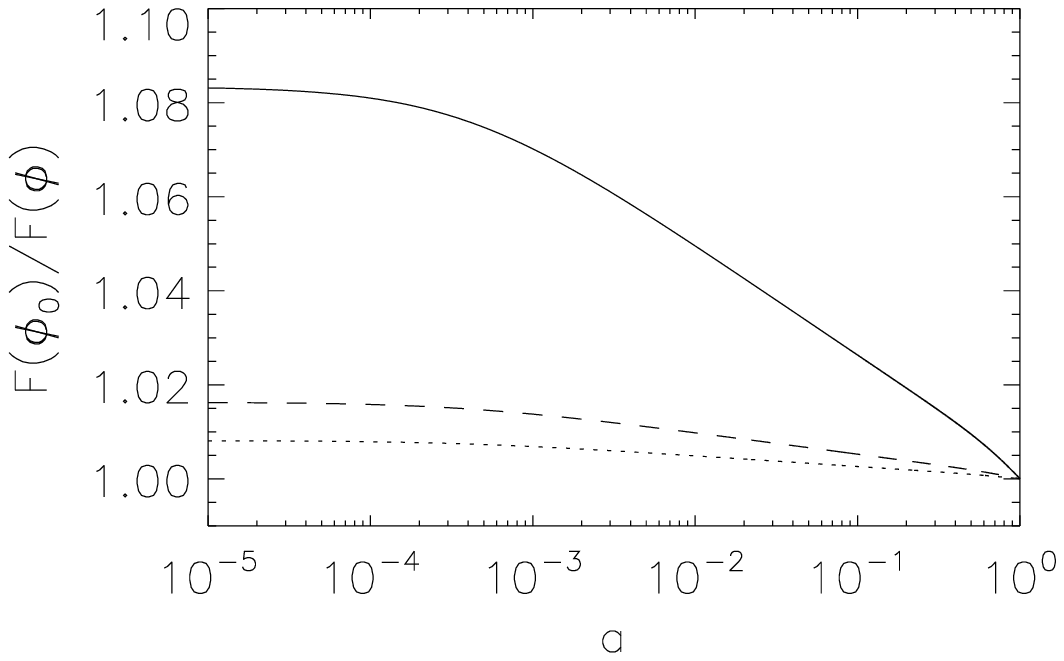}
\includegraphics[width=5cm]{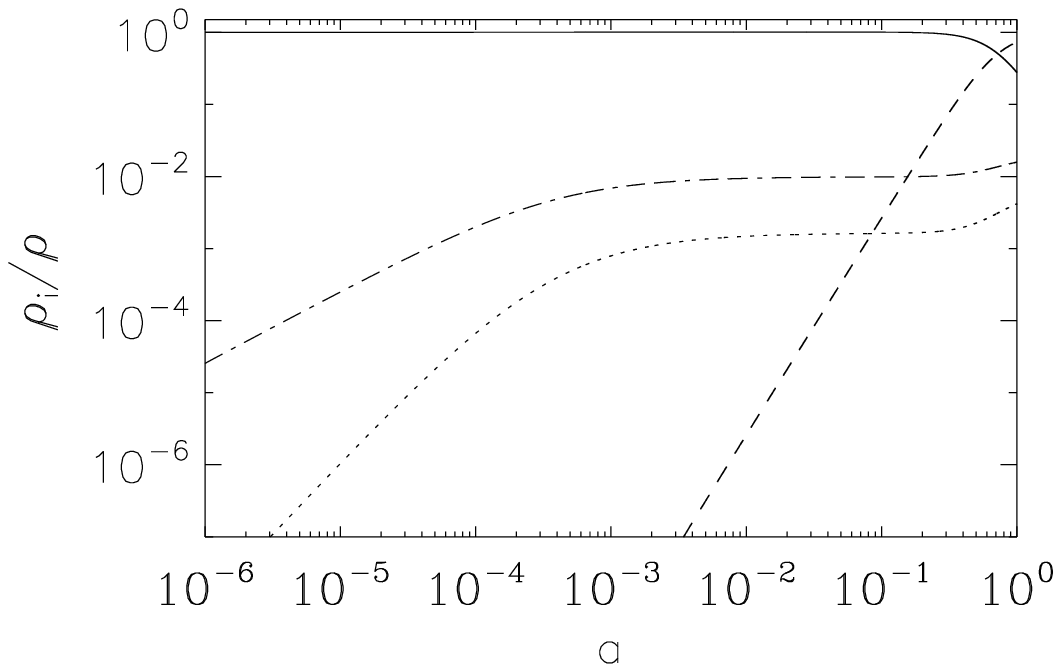}
\includegraphics[width=5cm]{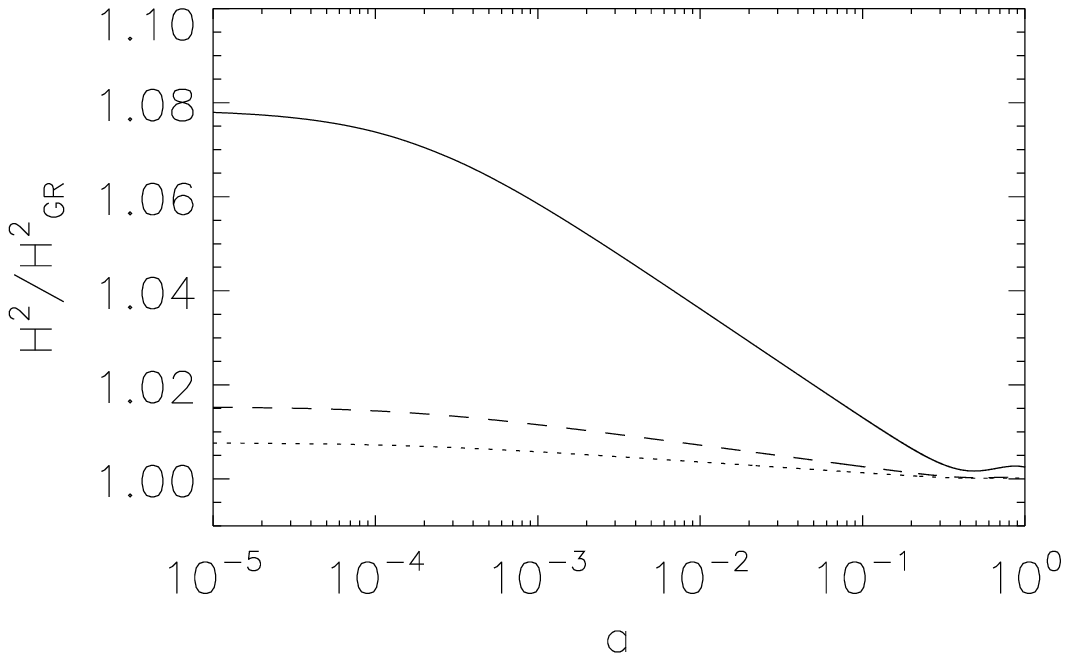}
\caption{Left and right panel: evolution of the gravitational coupling $1/F$, normalized to the present value, and of the squared Hubble factor, normalized to a GR model with the same cosmological parameters. We show curves for $\omega_{\rm JBD}$ = 100 (solid line), 500 (dashed line), 1000 (dotted line). Center panel: relative contributions of the four terms in the r.h.s. of Eq. (\ref{eq:h}): fluid part (solid line), kinetic term (dotted line), constant potential (dashed line), $\dot{F}$ term, plotted with opposite sign (dashed-dotted-line); trajectories are shown for a fiducial model with $\omega_{\rm JBD}$ = 100.}
\label{fig:trajectory}
\end{figure}

To study the phenomenology of the JBD model and its differences with GR, we start by comparing cosmological observables for the two models  assuming that they have the same cosmological parameters other than $\omega_{\rm JBD}$. The GR $\Lambda$CDM model is specified by 6 parameters: $\omega_b=\Omega_b h^2$, $\omega_{CDM}=\Omega_{CDM}h^2$, $n_s$, $\tau$, $A_s$, $h$; physical density of baryons, physical density of cold dark matter, primordial power spectrum spectra slope, optical depth to the last scattering surface, amplitude of the primordial perturbations and Hubble constant in units of $100$ km/s/Mpc. When comparing to a JBD model, we assign the GR $\Lambda$CDM model a formal value of $\omega_{\rm JBD} = \infty$. \\
It is useful to re-define the JBD field in the following way:
 \be
 \phi^2 = \frac{\omega_{\rm JBD}{\bf \Phi}}{2\pi}; \qquad \xi = \frac{1}{4\omega_{\rm JBD}}. 
\ee
With such notation the field  $\phi$ is now a canonical scalar field, non-minimally coupled to the Ricci scalar, and the Lagrangian of the JBD model reads
\be
\label{eq:lbd}
{\cal L_{\rm JBD}} = \frac{1}{2}\xi\phi^2 R - \frac{1}{2}\p_\mu \phi \p^\mu \phi - V(\phi)+ {\cal L}_{\rm fluid}.
\ee
For comparison, the Lagrangian for GR with a Cosmological Constant term is:
\be
\label{eq:lgr}
{\cal L_{\rm GR}} = \frac{1}{16\pi G} R - \Lambda + {\cal L}_{\rm fluid}.
\ee
To isolate the effect of the gravity modifications alone, we set the potential to be constant, mimicking a $\Lambda$ term, throughout all of our analysis. This corresponds to the "worst-case scenario" in terms of detectability, since in this case Eqs. (\ref{eq:lbd}) and (\ref{eq:lgr}) only differ in the gravity sector, and late-time acceleration is driven by the same mechanism. \\
The equations of motion for the field $\phi$ are given by
\be
{H}^2 + 2 {H} \frac{\dot{\phi}}{\phi} = \frac{2}{3}\omega_{\rm JBD}\left(\frac{\dot{\phi}}{\phi}\right)^2 + \frac{4}{3}\frac{\omega_{\rm JBD}}{\phi^2}\rho \,;
\ee
\be
\frac{\ddot{\phi}}{\phi} + \left(\frac{\dot{\phi}}{\phi}\right)^2 + 3 {H} \frac{\dot{\phi}}{\phi} = \frac{2 \omega_{\rm JBD}}{2 \omega_{\rm JBD}+3}\frac{\rho - 3 p}{\phi^2}\,.
\ee
Initial conditions for the evolution of the field can be set requiring that its present value reproduces the strength of gravity observed in Cavendish-type experiments in the local universe \cite{Will}:
\be
\label{eq:incond}
\phi_0^2 = \frac{\omega_{\rm JBD}}{2\pi G}\,\frac{2 \,\omega_{\rm JBD} + 4}{2 \,\omega_{\rm JBD}+3}.
\ee
The above equation is ensuring that at the present time the gravitational coupling $\xi \phi^2 \rightarrow 1/(8\pi G)$, so that GR is the late-time limit of theory for any value of $\omega_{\rm JBD}$ \footnote{There is indeed a higher order correction to the quoted limit, known as Cavendish correction, expressed by the second term on the right hand side of Eq. (\ref{eq:incond}). We take it into account in our numerical codes, but we will generally say, slightly inappropriately, that the theory recovers GR at the present time.}.\\
The other initial condition is $\dot{\phi}_{\rm beg} = 0$, since the trajectory of the field during the radiation era is constant, as known from analytical solutions in this regime \cite{Nariai:1969vh,1973Ap&SS..22..231G}. \\ 
With the notation $\xi \phi^2$ = $F(\phi)$, we can now explicitly compare the evolution of the Hubble factor for the JBD model and for  GR:
\begin{eqnarray}
\label{eq:h}
{H}^2_{\rm JBD}& =& \frac{1}{3 F}\left[ \rho_{\rm fluid} + \frac{1}{8 \pi G}\left(\frac{1}{2}\dot{\phi}^2 + a^2\,V - 3 {H} \dot{F}\right)\right]; \nonumber \\
\quad {H}^2_{\rm GR} &=& \frac{8 \pi G}{3} \rho_{\rm fluid} + \frac{\Lambda}{3}.
\end{eqnarray}
The first deviation from GR is expressed by the $1/3F$ term, which acts like an effective time-varying gravitational constant $G_{\rm eff}=1/(8\pi F)$. Its evolution is shown in the first panel of Fig. \ref{fig:trajectory}, for fiducial models with $\omega_{\rm JBD} = 100$, 500, 1000. We have normalized the curves to their present value, which also correspond to the standard GR gravitational constant since the theory approaches GR at late times.  It is a monotonic, decreasing function of time, implying that the gravitational force was stronger in the past. Its trajectory stays close to the initial value in the radiation and early matter domination era: for such redshifts the difference between GR and JBD is largest. For larger values of $\omega_{\rm JBD}$ the curves are closer to the GR value and their time variation is less significant. \\
The second difference is caused by the additional terms in the right hand side of Eq. (\ref{eq:h}). For the case $\omega_{\rm JBD} = 100$, we plot their relative contributions to the total energy density in the middle panel of Fig. \ref{fig:trajectory}.
The energy density is governed by the fluid-like components (matter and radiation) up to late times, when the effective cosmological constant (constant potential) takes over: this is similar to what happens in GR. The only difference is a slight shift due to the term $3 H\dot{F}/a^2$, which has a relative weight of a few per thousand up to one per cent for this value of $\omega_{\rm JBD}$, while the kinetic term is at least one order of magnitude smaller. \\
We conclude that the largest impact of these modifications of gravity is caused by the change in the effective gravitational constant, which is proportional to $1/F(\phi)$, and most relevant at early times. \\
Finally, the right panel of Fig. \ref{fig:trajectory} we show  the ratio ${H}^2_{\rm JBD}/{H}^2_{\rm GR}$; the different evolution of $H^2(z)$ encloses both the two effects described above. 

\section{Interpretation of the JBD field: early-time effects}
\label{early}
We now proceed to understand how evidence of non-GR gravity can be discovered, without assuming any {\it a priori} knowledge of the gravity theory.   \\ 
A key element in this respect, that has been extensively recognized in literature, is the fact that modifying gravity affects both the background evolution and the linear perturbations growth. n particular, in GR 
the redshift dependence of the matter density perturbation ($\delta_m \equiv \delta\rho_m/\rho_m$, where $\rho_m$ denotes the matter density) can be exactly predicted for a given  expansion history.
Thus in general, if one could measure $H(z)$ and $\delta_m$(z) at the same time and infinitely well, any modification of the underlying gravity theory would manifest as an incongruence between the prediction of GR and the actual measurement \cite{Lue1,Linder:2005in,Koyama:2005kd,Knox_Song_2,Upadhye,Ishak_Spergel,Huterer:2006mv,Kunz:2006ca,Bludman:2007kg,Wang:2007fsa}. \\ 
 However, modifications to the redshift evolution of $\delta_m$ in JBD result in maximum part from the different strength of the gravitational field, which is encoded, as seen in the previous Section, in the function $1/F$. Such function is significantly different from the GR case only at high redshift, since by construction the model tends to GR at late times.  One would need to accurately measure the matter density perturbations at early times in order to distinguish between the two cases. Unfortunately, at such epoch  direct measurements of $\delta_m(z)$ are expected to be extremely challenging. (More details on the evolution of $\delta_m$ in the JBD model can be found in Appendix A).  \\
We conclude that fixing the expansion history for such models also corresponds to have very similar perturbations growth rate at redshifts relevant for structure formation, so that the idea of detecting deviations from GR by means of these two observables cannot be applied in the present case. \\
\section{Early times vs late times observables} 
\label{earlylate}
 Although JBD modifications to GR do not manifest as incongruencies between the expansion history and the growth of
perturbations, a similar approach can still be used. We propose to use early and late time observables. 

The key element of this method resides in the fact that by construction our JBD models resemble GR in the local universe. Thus,  a standard GR, $\Lambda$CDM model  will be a good fit to low redshift data  and will recover the correct underlying  cosmological parameters whether or not the ``true'' theory is JBD or GR. Conversely, observations at high redshift would be sensitive to the modified gravitational coupling, and give different best fits for the same parameters, if gravity is JBD ($\omega_{JBD}<\infty$), but  agree with the  low redshift parameter fit if gravity is GR. 

Although we present and develop the method in the context of the JBD theory, the power of such method is that no previous knowledge of the model is assumed; the only requirement is to use different datasets separately according to their ``early'' or ``late'' time nature. Moreover, even if we use the JBD model as a working example, and will give limits on the JBD parameter, such inconsistencies are indeed a general indication of deviation from GR at some level.
 
 \subsection{Getting quantitative}
To quantify  how small deviations from GR  can be observed through the method described above, we will start by assuming that the ``true'' Universe is described by a JBD theory with a given value of $\omega_{\rm JBD}$. We will  then  generate a typical early-time observable, associate to it expected observational uncertainties, and   fit it with a GR $\Lambda$CDM model. This gives an estimate of the best fit  values of the recovered cosmological parameters, as well as of errorbars. For all the models which lie within 1$\sigma$ from the best fit model, we then make predictions for the value of late-time observables, again assuming GR. We then compute the late-time observable in the "true" JBD model and its associated expected observational uncertainties. 
We obtain two different confidence regions for the same observables: the first is predicted from an early-time observation and the (wrong) assumption of standard gravity, while the second is the confidence region allowed by observations. If the difference has enough statistical significance, it means that observations are able to discriminate between the two cases, JBD and Einsteinian gravity, for that particular value of $\omega_{\rm JBD}$.

\subsubsection{Early time observable}
For  the early-time  observable we choose the Cosmic Microwave Background (CMB),  as it can be measured  by the Planck experiment \cite{planck}. We use the Planck specifications as described in the Planck Blue Book \cite{planck_blue}. \\
 Let us note that, although not an object of study of the present work, another possible early time observable would be the the amount of baryons from Big Bang Nucleosynthesis, which is modified in JBD scenarios with respect to GR (e. g. \cite{Serna:1995tr,Santiago:1997mu}). The most stringent limit obtained on $\omega_{\rm JBD}$ from BBN constraints is $\omega_{\rm JBD}$ $\ge$ 32 \cite{Damour:1998ae}, confirmed by the analysis in \cite{Coc:2006rt}. Such limit improves significantly using the more recent, tighter constraints on the abundance of primordial deuterium to hydrogen ratio from \cite{Kirkman:2003uv,Crighton:2004aj}. Assuming that the current upper limit on D/H can be cast as $D/H < 4\times10^{-5}$, the limit on $\omega_{\rm JBD}$ tightens to $\omega_{\rm JBD}$ $\ge$ 90. 

We consider  several representative values for $\omega_{JBD}$: $100$, $200$, $500$, $1000$. This range is motivated as follows. As for the lower bound, \cite{Acquaviva:2004ti} showed that  the fitted values of cosmological parameters for a JBD case with $\omega_{\rm JBD} = 70$  are already ruled out by data. For the upper bound, a Fisher matrix analysis for a JBD model and the Planck experiment forecasts a $1-\sigma$ detection threshold of the $\omega_{\rm JBD}$ parameters between 1000 and 2000 \cite{Chen_Kam}. While the Fisher matrix approach already assumes knowledge of the underlying model (and thus give more stringent constraints than model independent methods), the method presented here relies in the additional statistical power of late time observables.  

\begin{figure}
\centering
\vspace{0.5cm}
\includegraphics[width=5cm]{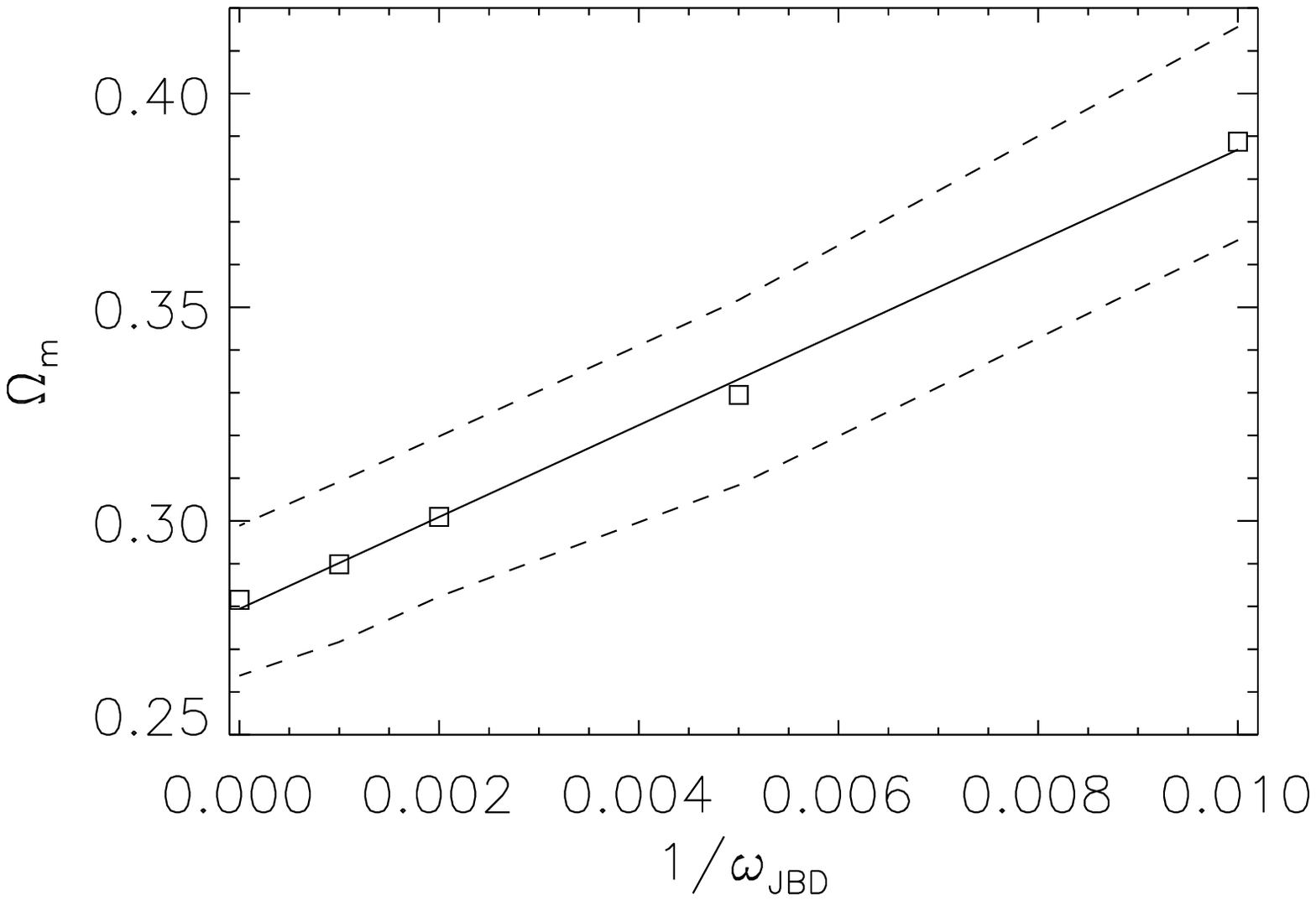}
\includegraphics[width=5cm]{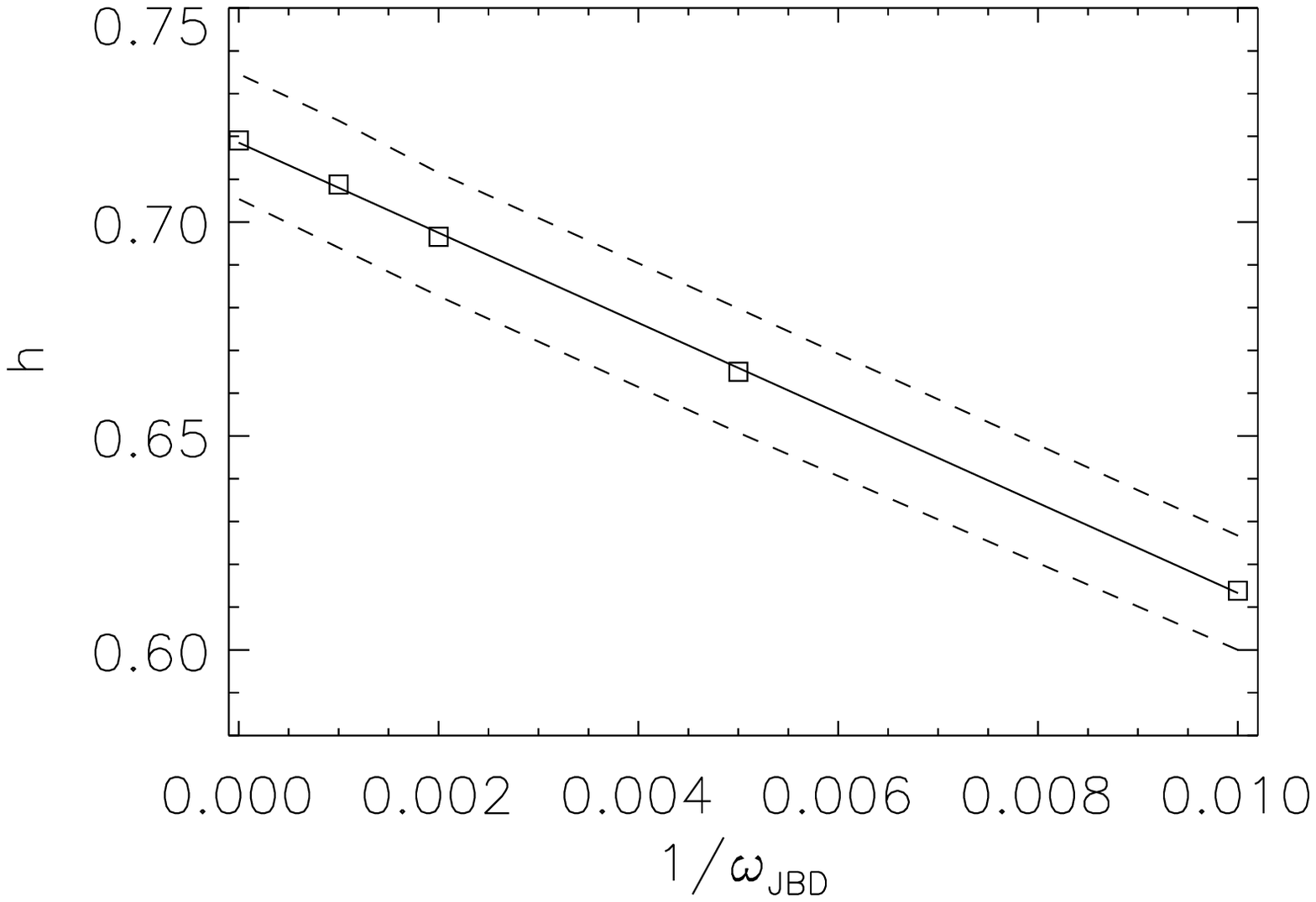}
\includegraphics[width=5cm]{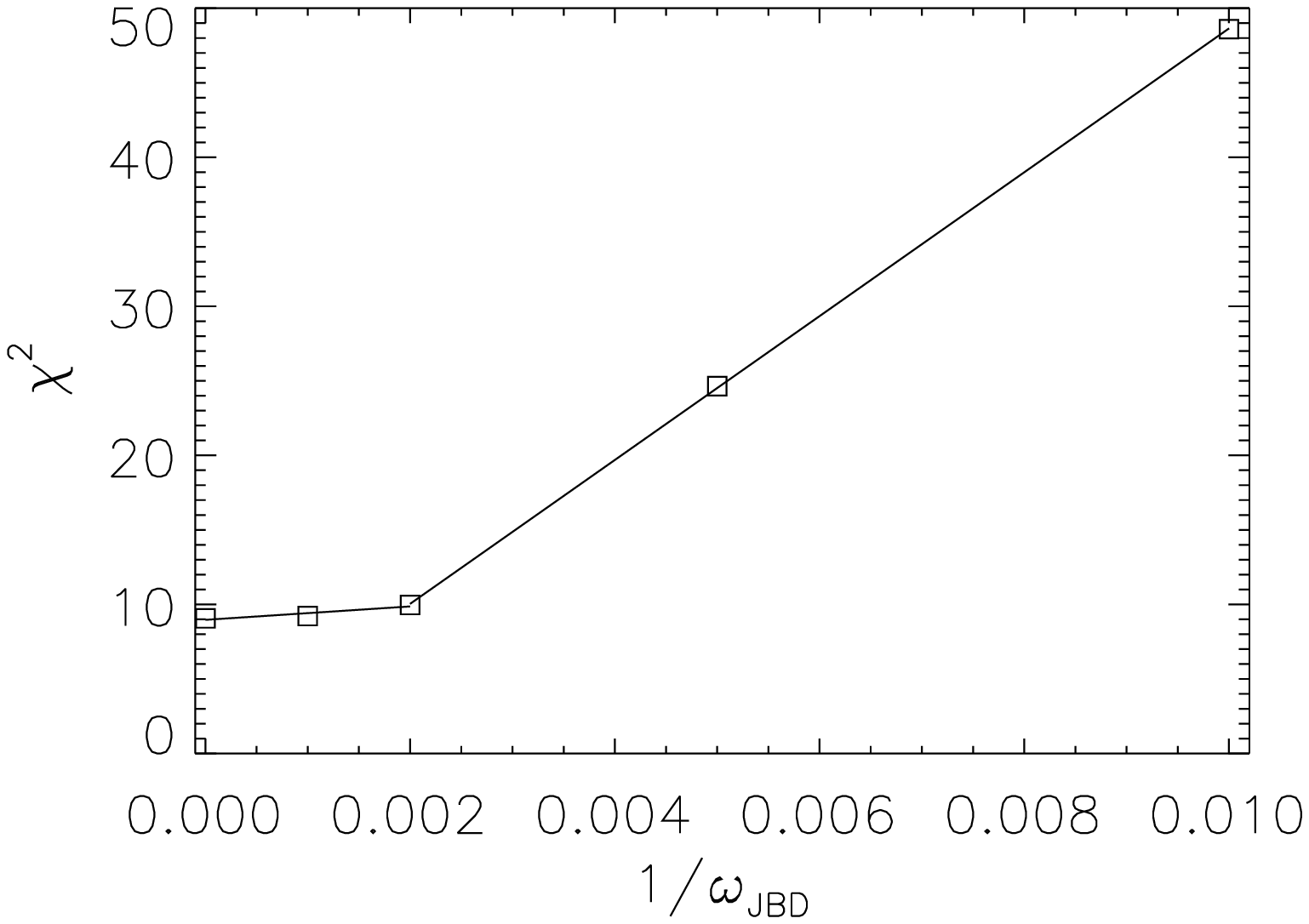}
\caption{Linear fit of $1/\omega_{\rm JBD}$ versus $\Omega_m$, h and $\chi^2$ as recovered from a fit to a GR $\Lambda$CDM model. Data points are shown for $\omega_{\rm JBD}$ = 100, 200, 500, 1000 and $\infty$; the last case corresponds to GR.}
\label{fig:fit_shift}
\end{figure}

For each of  the reference JBD models, we compute the ``true'' CMB temperature, EE polarization, and their cross-correlation power spectra with the DEfast code \cite{Perrotta:1998vf,ext_quint,track_extquint}, originally based on the CMBfast package \cite{SZ96}. We find the best fit GR $\Lambda$CDM parameters and their confidence regions by  running  Markov Chain Monte Carlo chains, using the COSMOMC code \cite{cosmomc}, which uses CAMB \cite{camb}.
  One of our concerns was that  small numerical differences between the two codes could give rise to systematics effects in the recovered parameters. In fact, even running the two codes with settings as close as possible, some numerical differences in the CMB power spectra remain. We  find that  such difference is independent of the cosmological model, and  we correct for it, as described in detail in Appendix B. 
In Table \ref{tab:chains} we show the best fit cosmological parameters obtained fitting CMB data for various $\omega_{\rm JBD}$; we report uncertainties at the $2 \sigma$ level. 
\begin{table}[htbp!]
\small
\begin{center}
\vspace*{0.3cm}
\caption{Best fit and $2 \sigma$ confidence levels of cosmological parameters from Planck CMB forecasts, for JBD models with $\omega_{JBD}$ = 100, 200, 500, 1000, $\infty$, fitted with a $\Lambda$CDM model. ${\cal A}\equiv \ln (10^{10}A_s)$.}
\vspace*{0.3cm}
\begin{tabular}{|l|c|c|c|c|c|c|c|}
\hline 
&  input & $\omega_{\rm JBD} = 100$& $ \omega_{\rm JBD}=200$ &
  $\omega_{\rm JBD}=500$ & $\omega_{\rm JBD} = 1000$ & $\omega_{\rm JBD} = \infty$\\
& JBD model &$\Lambda$CDM  fit   & $\Lambda$CDM  fit &  
$\Lambda$CDM  fit   &$\Lambda$CDM  fit &$\Lambda$CDM  fit  \\
\hline
 $ { \omega_b }$ &  $ 0.022 $ & $0.0215^{\, 0.0218}_{\, 0.0212}$ & 
 $0.0217^{\, 0.022}_{\,0.0214}$ & $0.0219^{\,0.0223}_{\,0.0216}$ & $0.022^{\,0.0223}_{\,0.0217}$  & $0.022^{\,0.0223}_{\,0.0217}$ \\
 ${ \omega_{\rm CDM} }$ &  $0.1232$ & $0.1240^{\,0.1275}_{\,0.1206} $ & $0.1250^{\, 0.1284}_{\, 0.1219}$ & $0.1241^{\,0.1273}_{\,0.1206}$ & $0.1236^{\,0.1272}_{\,0.1201}$ & $0.1236^{\,0.1269}_{\,0.1202}$\\
 ${n_s} $ & $0.95$ & $0.9144^{\,0.9215}_{\,0.9053}$ & $0.9295^{\,0.9373}_{\,0.9159}$ &
 $0.9448^{\,0.9532}_{\,0.9365}$ & $0.9478^{\,0.9561}_{\,0.9395}$ & $0.9511^{\, 0.9588}_{\,0.9429}$ \\
 ${ \tau}$ & $0.09 $ & $0.0847^{\,0.0927}_{\,0.0745}$ & $0.0873^{\, 0.0988}_{0.0767}$ &
  $0.0900^{\,0.0986}_{\,0.0818}$ & $0.0905^{\,0.1022}_{\,0.0798} $  & $0.0915^{\,0.1023}_{\,0.0809}$\\
${\cal A}$ & $3.1355 $ & $3.0968^{\,3.1175}_{\, 3.0753} $ & $3.1112^{\, 3.1333}_{\, 3.0991}$ &
$ 3.1349^{\,3.1586}_{\,3.1124}$ & $3.1376^{\,3.1610}_{\, 3.1158}$ & $3.1411^{\,3.1637}_{\, 3.119}$ \\
 ${ h_0 }$ & $0.72 $ & $ 0.6138^{0.6267}_{0.6000} $ & $0.6650^{\,0.6798}_{\, 0.6508}$ & 
 $0.6965^{\,0.7115}_{\,0.6827}$ & $0.7088^{\,0.7237}_{\,0.6940}$ & $0.7191^{\,0.7346}_{\,0.7054}$\\
 ${ \Omega_m }$ & $0.28 $ & $0.3887^{\, 0.4156}_{\,0.3657}  $ & $0.3295^{\,0.3517}_{\, 0.3084}$ &
   $0.3009^{0.3198}_{0.2823}$ & $0.2899^{\,0.3092}_{\, 0.2717}$ &$0.2815^{\,0.2989}_{\,0.2638}$  \\
{$\chi^2$} &  0  &  48.6 & 24.64 & 
9.96 & 9.22 & 9.06 \\
 \hline
\label{tab:chains}
\end{tabular}
\end{center}
\normalsize
\end{table}
For reference, we also show how well we recover the cosmological parameters for a $\Lambda$CDM model, formally identified by $\omega_{\rm JBD} = \infty$. \\
Some of the parameters are quite insensitive to the presence of the JBD field, and their predicted values coincide with those of the ``true'' JBD model within errorbars, even for values of $\omega_{\rm JBD}$ as small as 100. This is the case for $\tau$ and the primordial amplitude of the perturbations, $A_s$. Interestingly, the combination $\omega_{\rm m} = \Omega_{\rm CDM} h^2 + \Omega_b h^2$ is also recovered quite well. In fact, such combination is mainly constrained by the distance to last scattering, which is in large part integrated over look-back times where the field is unimportant.
However, $h$ and $\Omega_m$ are respectively under- and over-estimated. In fact, the size of the sound horizon at recombination is significantly smaller in a JBD scenario than in a $\Lambda$CDM (e.g. \cite{Riazuelo:2001mg}), since it is only affected by pre-recombination physics, where the relative weight of the field component in the total density is significant. As we have seen in Sec. 3, gravity in GR is weaker at early times  than in JBD and the JBD phenomenology is in part mimicked by an additional matter field. This moves the peaks towards smaller scales; since the distance to last scattering is constrained, such effect can be only accounted for enhancing the matter component, and lowering the Hubble factor. \\
Note that even for a value of $\omega_{\rm JBD}$ as large as 500,  the shift in $\Omega_m$ from its true value is as large as $7\%$, and that in $h$ is between 3 and 4 $\%$; furthermore, their ``true'' values lie outside the  $2-\sigma$ confidence contours obtained through Planck-quality CMB data. This is interesting because these parameters can be constrained well through late-time observables. \\
Finally, there is an overall shift of power towards smaller scales, corresponding to an underestimation of the primordial power spectrum index, $n_s$, also resulting from the change in the location of the peaks. \\
In the last entry of the table we report the $\chi^2$ for the best fit. The value of $\chi^2$ =9.06 for the GR case $\omega_{\rm JBD}=\infty$ arises from numerical differences between DEfast and CAMB. For reference, the value of the $\chi^2$ for the GR model if the same code, either DEfast or CAMB,is used both for generating the CMB spectra and for the fitting procedure is of order unity or smaller (see Appendix B for more details).\\
\begin{figure}
\includegraphics[width=7cm]{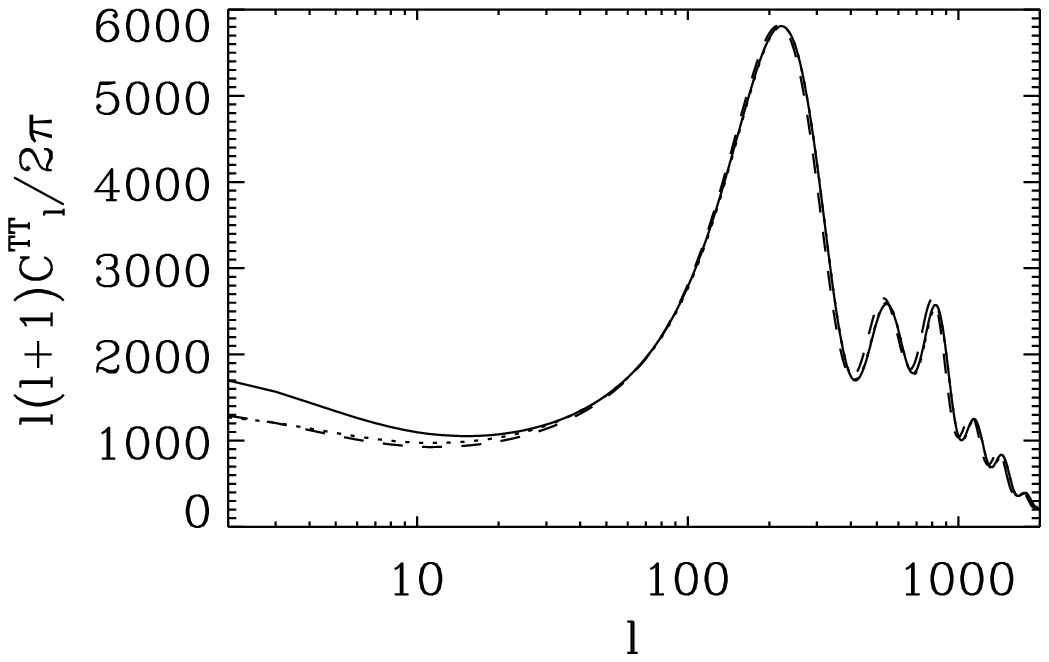}
\hspace{0.5cm}
\includegraphics[width=7cm]{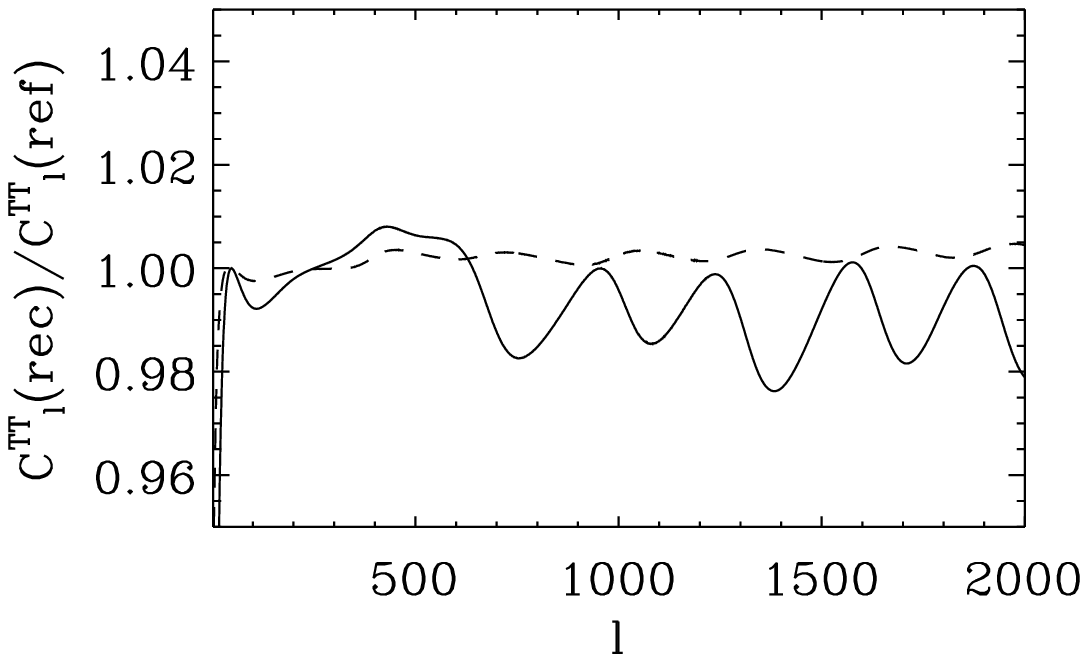}
\caption{Left panel: Temperature power spectrum of the reference JBD model with $\omega_{\rm JBD} = 100$ (solid line), compared to a GR $\Lambda$CDM model with same cosmological parameters (dashed line), and to the reconstructed GR  $\Lambda$ CDM model with parameters of Table \ref{tab:chains} (dotted line). Although the reconstructed model resembles the ``true'' JBD model much more closely, some differences are still present. Right panel: Ratio of the reconstructed temperature power spectrum and the input ``true'' JBD model, for $\omega_{\rm JBD} = 100$ (solid line), and $\omega_{\rm JBD} = 500$ (dashed line). The residual difference seen for $\omega_{\rm JBD}$ = 100 is reflected by the the high value of $\chi^2$ of the fit.} 
\label{fig:cls}
\end{figure}
In Fig. \ref{fig:cls} we show the temperature power spectrum for the reference JBD model with $\omega_{\rm JBD}$ = 100, for a $\Lambda$CDM one with the same parameters, and for the best-fit $\Lambda$CDM, with the parameters of Table \ref{tab:chains}. There is a residual disagreement between the spectrum obtained from the fitting procedure and the ``true'' JBD spectrum, reflected by the relatively poor value of the $\chi^2$ of the fit. In particular, differences are seen in the ISW amplitude and in the height of the first two peaks. The first is generated by the clustering properties of the JBD field, which cannot be reproduced with a smooth component; the second comes from the slight underestimation of the baryon content. The different amount of ISW could in principle be used in order to detect modifications to GR \cite{ext_quint}, but its signal-to-noise level is expected to be below the cosmic variance for the target values of $\omega_{\rm JBD}$ considered in this work. \\
Such differences becomes negligible as long as $\omega_{\rm JBD}$ is larger than 500, as can be seen in the right panel of Fig. \ref{fig:cls}. We show the ratio of the temperature power spectra of the GR $\Lambda$CDM model coming from the fit and of the true JBD model input, for $\omega_{\rm JBD}$ = 100, 500 respectively. 
We conclude that for values of $\omega_{\rm JBD} $ smaller than or equal to 500, high values of the $\chi^2$ are already a sign that the fit is done using the wrong model. More sophisticated techniques of Bayesian analysis may be enough to show that more cosmological parameters are needed, \eg \cite{verde_heavens,Liddle04,Trotta05,Gordon:2007xm} . However, in the regime where $\omega_{\rm JBD} $ is larger than 500, the fit is almost as good as the one for the GR case. For such values, one can not rely on the CMB alone, coupled to goodness-of-fit techniques; the consistency between early and late time observables needs to be used. This paper focuses on this regime. \\
One interesting feature of the observed shift in the cosmological parameters, as reconstructed by the ``wrong'' fit, is that it scales almost linearly with $1/\omega_{\rm JBD}$, at least for the most sensitive $\Omega_m$ and $h$. The value of the $\chi^2$ also follows a similar pattern, but the fit is different for values of $\omega_{\rm JBD} \leq$ and $>$ 500, since the $\chi^2$ is almost constant above this value. We show such linear fits, of the form of ${\rm par}_{\rm fit} = a + b\times(1/\omega_{\rm JBD})$, in Fig. \ref{fig:fit_shift}; for reference, we also plot the $2 \sigma$ limits coming from the fit. Coefficients of the fit are $a$=(0.294, 0.719, 0.385) and $b$=(10.52,-10.75, 4286) for $\Omega_m$, $h$ and $\chi^2(\omega_{\rm JBD} < 500)$ respectively; for larger values of $\omega_{\rm JBD}$ the $\chi^2$ is fitted by $a_{\rm plateau}$ = 8.96; $b_{\rm plateau}$ =  450.3. This property is useful because it allows to quantify in a simple way the effect of the JBD field on the cosmological parameters. In particular, it can be used to estimate which target $\omega_{\rm JBD}$ experiments could detect, starting from the attainable precision on the cosmological parameters. 
 
\subsubsection{Late time observables}
We use the Monte Carlo simulations described above and  select all models within $68.3$\% confidence level from  the best fit $\Lambda$CDM model; for these models we extrapolate the values of late-time observables. 
The late time observables we consider are: the distance modulus as seen through Supernov{\ae} observations, $\mu(z)$; the angular diameter distance $d_A(z)$ and the Hubble factor $H(z)$, by means of the Baryon Acoustic Oscillations; and the power spectrum of convergence $P_\kappa$, with weak lensing surveys. 
We compute the same late-time observables  for the JBD fiducial models  and their respective observational errors for different experimental set ups. We concentrate here on the possibility of constraining values of $\omega_{\rm JBD} \geq$ 500, for which, as we said, we cannot use information about the quality of the fit, postponing the analysis of lower values of $\omega_{\rm JBD}$ to future work \cite{work_in_prep}. we therefore consider, at first, datasets that will be available around 2010 ("2010"); and second, datasets likely to be available a decade later ("2020"). \\
We evaluate observational errors, for each observable, as follows; 
specifics of the experimental setups considered can be found in Tables \ref{tab:sne}, \ref{tab:BAO}, \ref{tab:WL}. \\

{\bf SNe.} For Supernov{\ae} type Ia, used to
constrain the luminosity distance modulus, $\mu$(z):
\be
\mu(z) = 5 \log_{10}(d_L(z)) + 25; \quad d_L(z)=(1+z)\int^0_z \frac{1}{H(z)}dz\,,
\ee 
we evaluate statistical errors following the treatment in \cite{DETF}. Errors per each supernova are obtained adding in quadrature the uncertainty of the corrected apparent magnitudes due to the variation in the properties of SNe, $\sigma_D$, and the measurement uncertainty, $\sigma_m$. We neglect here the change in shape of the light curve of the Supernov{\ae} due to the variation in time of the gravitational constant \cite{Riazuelo:2001mg}, since by construction GR is the correct description of the Universe at late times.
Systematic errors are estimated according to the prescription in \cite{linder_huterer}, as
\be
\label{eq:syst}
\sigma_s(z) = A\,(1.7/z_{\rm max})(1+z)/2.7\,,
\ee
where A is 0.05 for ground-based surveys, and 0.02 for space-based surveys, and $z_{\rm max}$ is the depth of the survey. \\
For the "2010" scenario we consider two cases: the Dark Energy Task Force Stage II for Supernov{\ae} (DETFII), as in Table 3.2 of \cite{DETF}, and the combination of the on-going ESSENCE \cite{ESSENCE}, SNLS \cite{SNLS,SNLS2} and SDSSII \cite{SDSSII} surveys.
 We refer to such configuration as ``ONGOING'', and we assume a redshift distribution similar to the DETFII case (other than the local sample).
 For the "2020" scenario we consider a survey such as LSST or SNAP, corresponding to Stage IV of Ref. \cite{DETF}. \\

{\bf BAOs.} Baryon Acoustic Oscillations can be used to constrain separately the expansion history, $H(z)$-from the line-of-sight clustering, and the comoving angular diameter distance, $d_A(z) = a\,d_L(z)$, if spectroscopic surveys are used. 
We follow \cite{Blake:2005jd} to forecast errors for both spectroscopic and photometric surveys;
we always assume that systematic errors  are below the statistical errors \cite{Seo:2007ns}. 
In the case of the BAO, the time line of ``2010'' is spread out over a few years.  The first setup we consider is the SDSS LRG sample \cite{Eisenstein:2005su}; 
an improved version of the same survey (LRG BOSS) \cite{BOSS}, 
and the PAU-BAO survey \cite{PAU-BAO}.
 For the "2020" scenario we consider a survey like ADEPT \cite{adept}.  \\

{\bf Weak lensing.} Future weak lensing observations are used in order to constrain the power spectrum of the shear in multipole space, $P_\kappa$. In the Limber approximation it is written as \cite{BS2001}:
\be
P_{\kappa}(l) = \frac{9}{4} H_0^4 \Omega_m^2 \int^\chi_{H_0} \frac{g^2(\chi)}{a^2(\chi)} P(\frac{l}{\chi},\chi) d\chi,
\ee
where
\be
g(\chi)=\int^{\chi_H}_\chi n(\chi') \frac{\chi' - \chi}{\chi'}d\chi'
\ee
and $n(z)$ is the normalized source distribution, which we assume to be of the form $n(z) = (z/z_0)^2 \times e^{(-(z/z_0)^{3/2})}$. We compute the growth factor and matter transfer function using the analytical formulas of \cite{EisensteinHu}; we use the same approximation for the ``true'' JBD model matter power spectrum. We obtain the nonlinear matter power spectrum correction using the prescription of Peacock and Dodds \cite{PD96}. We however discard multipoles beyond $l = 1500$, in order to avoid errors coming from the uncertain nonlinear galaxy evolution and from baryonic physics \cite{Jing:2005gm,Zhan:2004wq}.  Errors on the shear power spectrum are obtained as
\be
\delta P_{\kappa}(l) = \sqrt{\frac{2}{(2 l + 1)f_{\rm sky}}}\left(P_\kappa(l) + \frac{\langle \gamma_{\rm int}^2 \rangle}{\tilde{n}}\right) + \sigma_{\rm deg};
\ee
where $\gamma_{\rm int}$ is the rms ellipticity per galaxy, assumed to be 0.16, ${\tilde{n}}$ is the mean number of observed galaxies per square arcminute, and $\sigma_{\rm deg}$ is the error coming from the photo-z errors in the determination of redshifts of sources. We estimate this error to be $1.6 \times \sigma_z$, where $\sigma_z$ is the photo-z error of a given survey, and we have used the analytic fitting formula for the shear variance (\eg \cite{jain_seljak}). \\
For the ``2010'' weak lensing scenario, we consider a configuration like DES \cite{des}, 
and one like Pan-STARRS\cite{panstarrs},  while for the ``2020'' scenario we assume a survey like DUNE \cite{DUNE,Refregier:2006vt} or LSST \cite{LSST}.

\subsection{Results: forecasts for planned and future experiments}
We report specifics of the experimental setups considered and results for the statistical significance of detection of $\omega_{\rm JBD}$ = 500 and 1000 in Tables \ref{tab:sne}, \ref{tab:BAO}, \ref{tab:WL}. For these two values of $\omega_{\rm JBD}$, we compute the quantities $\Delta d_L$, $\Delta d_A$, $\Delta H$ and $\Delta P_\kappa $, defined as in 
\be
\Delta {\rm obs} = {\rm obs( chain)} - {\rm obs(true)}.
\ee
For each observable, ``chain'' refers to the value obtained using the cosmological parameters coming from the Monte Carlo chain in order to compute late time observables, and ``true'' is the observable for the fiducial JBD model. If gravity was GR, all the quantities above would be identically zero, because early and late time observables would be consistent with each other. Conversely, since we are assuming the the true Universe is a JBD, gravity modifications will manifest as a non-zero value for $\Delta d_L, \Delta d_A$ and $\Delta P_\kappa$. \\
We compute errors as follows, using $\Delta d_L$ as an example. For all the models in the $1 \sigma$ vicinity of the chain, $d_L ({\rm chain})$ will be larger than $d_L$(true), because of the shift in the cosmological parameters as recovered by the chain. The error on $d_L$(chain) is evaluated as the difference between $d_L$ for the best fit model and the minimum value of $d_L$ for models within $1\sigma$ from it; the error on $d_L$(true) is the experimental one around the true model, discussed in the previous Section.  The total error on $\Delta d_L$ is obtained adding in quadrature errors on $d_L$(chain) and on $d_L$(true). The statistical significance of a deviation of $\Delta d_L$ from zero is given by the $\Delta \chi^2$ of $\Delta d_L$, evaluated using this total error.

The value $\omega_{\rm JBD}$ = 500 is found to be within reach of all the next-generation experiments (with a little delay for the BAO projects with respect to the other probes). It is however interesting to compare how different setups can lead to different significance levels. For type Ia Supernov{\ae}, in the ``2010'' time range, we have considered two configurations, whose basic difference is the presence of $\simeq$ 350 more Supernova{\ae} in the local sample for the DETFII case \cite{DETF}. Such drastic reduction of statistical error in the first bin would be expected to increase the statistical significance of detection. However, for our adopted estimate of Eq. (\ref{eq:syst}) for the systematic errors in ground-based surveys, having more than $\simeq 100$ SNe per redshift bin of $\Delta z$ = 0.1 does not improve the signal-to-noise. This explain why the two surveys give similar performances. Both of them will be able to detect a JBD parameter $\omega_{\rm JBD}$ = 500 at the $3 \sigma$ level. As for the ``2020'' scenario, a survey like LSST or SNAP will gain information both from the richness of the local sample and the deepness of the whole survey. In this case, we found that a JBD model with $\omega_{\rm JBD}$ = 500 can be distinguished from a GR, $\Lambda$ CDM one with a significance of more than 7 $\sigma$, and a value of $\omega_{\rm JBD}$ = 1000 can be detected at the 3.3 $\sigma$ level.

For the BAO, we first consider constraints coming from both the comoving angular diameter distance, $d_A$, which we anticipate to carry most of the signal-to-noise, and $H(z)$. Errors are driven by the fraction of the total Universe volume covered by observations, so that a shallow ($z_{\rm max} <$ 0.5) survey, such as the first setup of SDSS LRG, does not provide any significant detection of $\omega_{\rm JBD}$ = 500. Going up to $z \simeq$ 0.75, as in the SDSS LRG BOSS configuration, is enough to get a first detection at the $2.3 \sigma$ level, and the PAU-BAO project, which will reach a redshift of $0.9$ and will cover $30\%$ more square degrees, is competitive with the ``2010'' SNe probes described above. For comparison, in order to get a $3 \sigma$ detection of $\omega_{\rm JBD}$ = 500, the shallow SDSS LRG survey should be able to observe over 30000 square degrees, more than 4 times its present sky coverage. In the ``2020'' experiments class, for a survey like ADEPT we forecast a detection of $\omega_{\rm JBD}$ = 500 at the 7.3 $\sigma$ level, and a detection of $\omega_{\rm JBD}$ = 1000 at 3.5 $\sigma$. For both values, this result is at the same level, or slightly better, than the corresponding SNe experiments planned on analogous time scale. \\
It is interesting to study what part of the signal-to-noise comes from information on $H(z)$ along the line-of-sight, as allowed by spectroscopic surveys like the ones we consider. 
We find that the $H(z)$ information  is  very useful for shallow surveys. In fact, with the usual notation, the difference 
\be 
\Delta H = H(z)({\rm chain}) - H(z)({\rm true})
\ee 
is larger at low redshifts and almost vanishes at redshift larger than 1. As a result, the significance of this measurement is comparable with that of $\Delta d_A(z)$ for the LRG BOSS, or the PAU-BAO. On the other hand, for a survey like ADEPT, where information is collected at high redshift, measurements of $H(z)$ do not substantially improve the performance of the experiment in detecting $\omega_{\rm JBD}$.  

For the weak lensing we forecast a significant detection of both $\omega_{\rm JBD}$ = 500 and 1000. In the case of $\omega_{\rm JBD}$ = 500, we obtain a $4.1 \sigma$ significance of detection with an experiment like DES, and the number increases to 6.0 $\sigma$ for a survey like Pan-STARRS, which should also be able to detect the value $\omega_{\rm JBD}$ = 1000 with a 3.6$\sigma$ significance. As for the ``2020'' scenario, a survey like DUNE or LSST will be able to reveal a value of $\omega_{\rm JBD}$ = 500 and $\omega_{\rm JBD}$ = 1000 with a 6.8 and 3.8 $\sigma$ significance, respectively. The fact that Pan-STARRS and DUNE/LSST give similar results is due to the fact that for such large sample of galaxies, the main error on $\Delta P_\kappa(l)$ is due to the half-width of the $1\sigma$ region within the best fit of the chain. The latter is only determined by the value of $\omega_{\rm JBD}$, and does not depend on the survey. These results from lensing can be understood if we consider the known (approximate) dependence of the shear variance on the parameters $\Omega_m$, $\Gamma$, and $\sigma_8$ (\eg \cite{jain_seljak}), and study how the shear power spectrum varies with the shift of cosmological parameters. Combining the different contributions, we found that there is a strong ($d\,log\, P(\kappa)/d\, log\, \Omega_m >$  1) dependence on the shift in $\Omega_m$, which is, as we have seen, the most sensitive parameter to the JBD field. Accordingly, the model reconstructed from the chain has $\simeq 10\%$ more power at all scales than the ``true'' JBD model for $\omega_{\rm JBD}$ = 500, and $ \simeq$ 5$\%$ more power for $\omega_{\rm JBD}$ = 1000; these differences are larger than those found in all the other observables. \\
However, as we already pointed out in the previous Section, we have only included the effect of some type of systematics. We have discarded modes above $l=1500$, possibly eliminating sources of errors related to nonlinear evolution and baryonic physics, and we have accounted for the uncertainties in the determination of the source redshift; nonetheless, other types of  systematic uncertainties, such as intrinsic alignment, PSF  correction etc., may remain, possibly with amplitude comparable to the signal.  For example, since the truly observable quantities are the correlation functions rather than the power spectrum $P_{\kappa}$, a further source of error is introduced in the mapping between the two \cite{Schneider:2002jd}. The fact that correlation functions cannot be observed over an infinite range of angles may bias the reconstruction of $P_{\kappa}$, which manifests as a fictitious oscillatory feature. The impact of such systematics is difficult to quantify extrapolating from the errorbars of the current, much smaller surveys; it may however contribute to degrade the efficiency of the weak lensing observations.

\begin{figure}
\includegraphics[width=7cm]{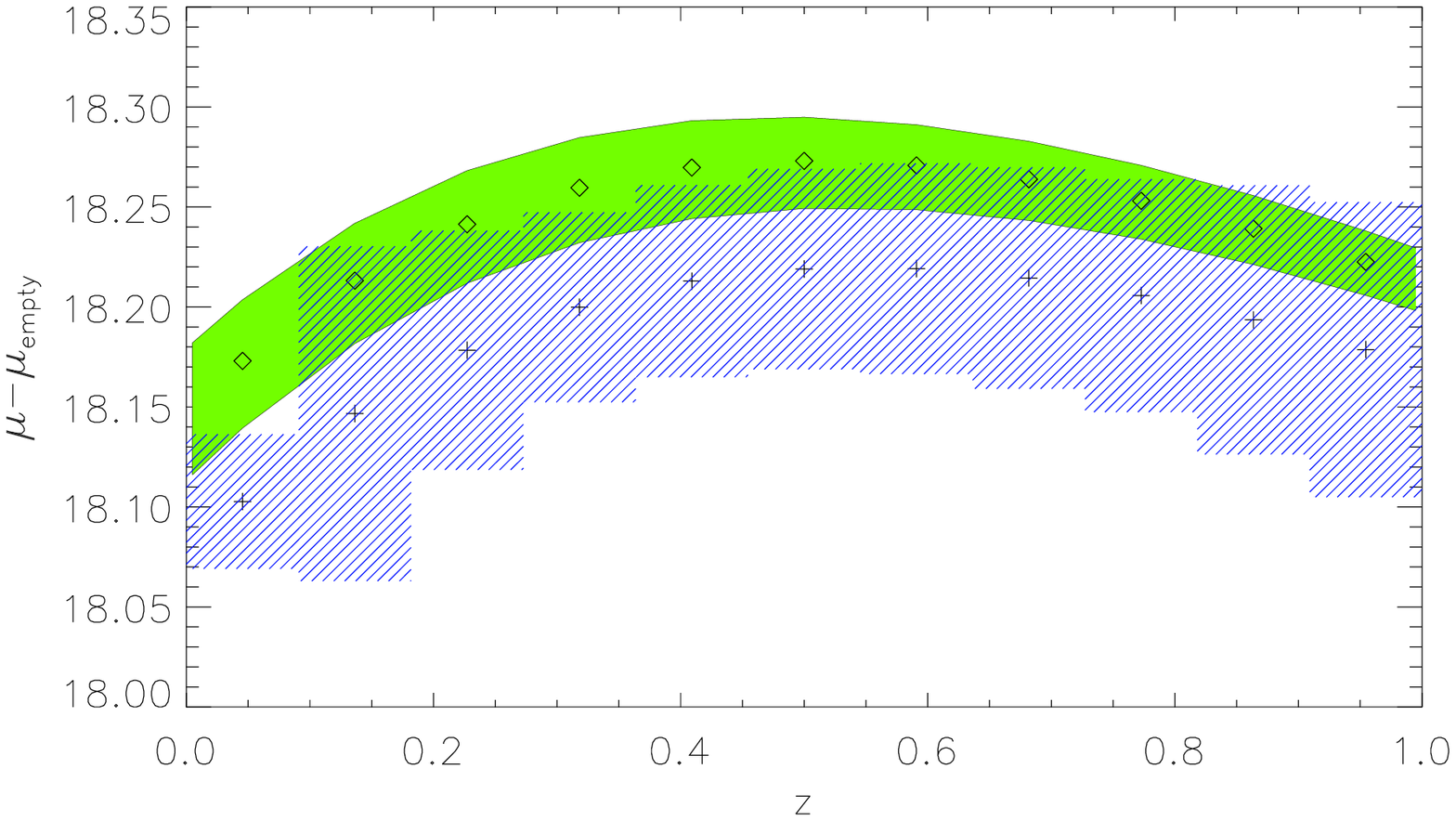}
\hspace{0.5cm}
\includegraphics[width=7cm]{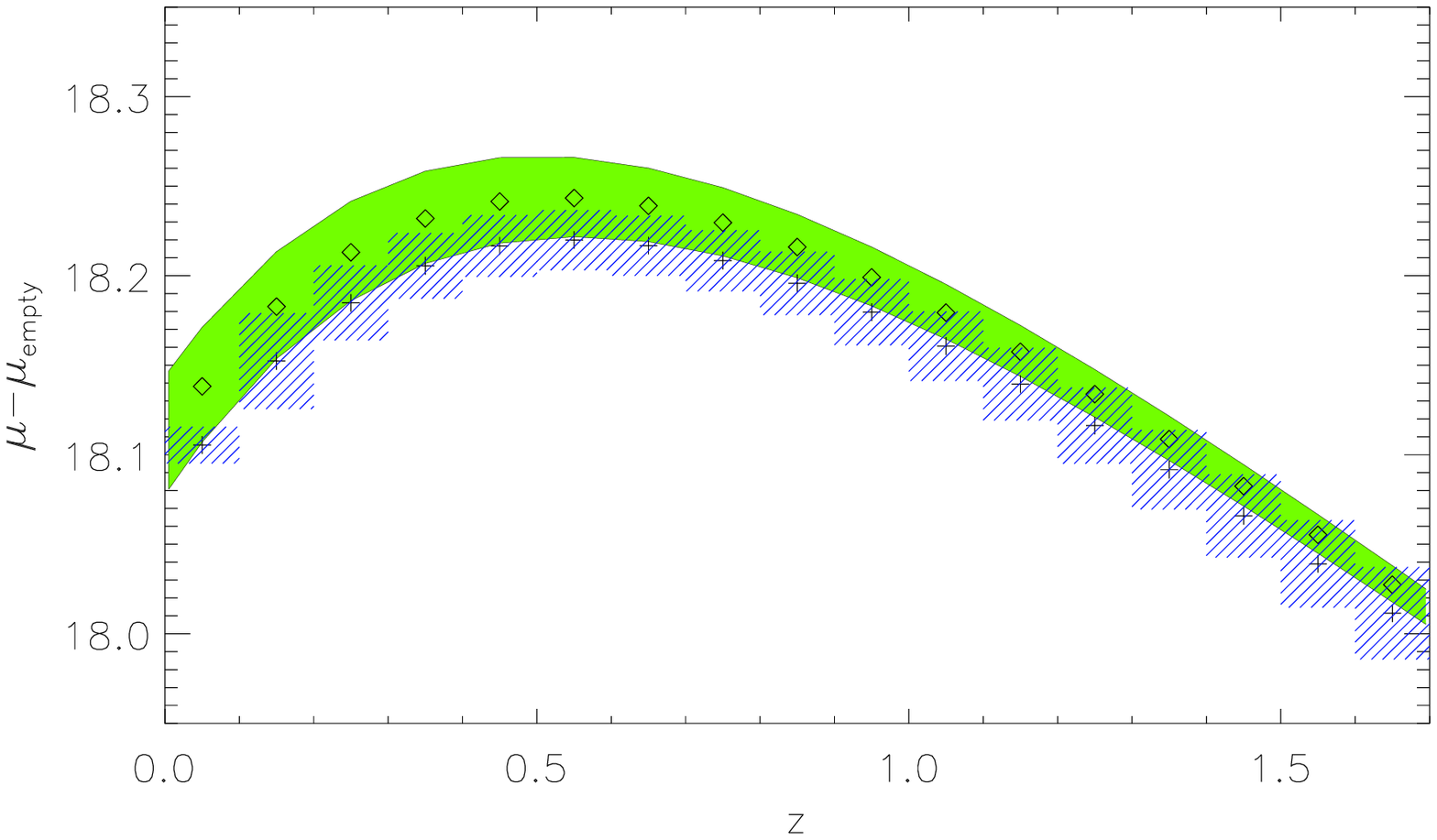}
\includegraphics[width=7cm]{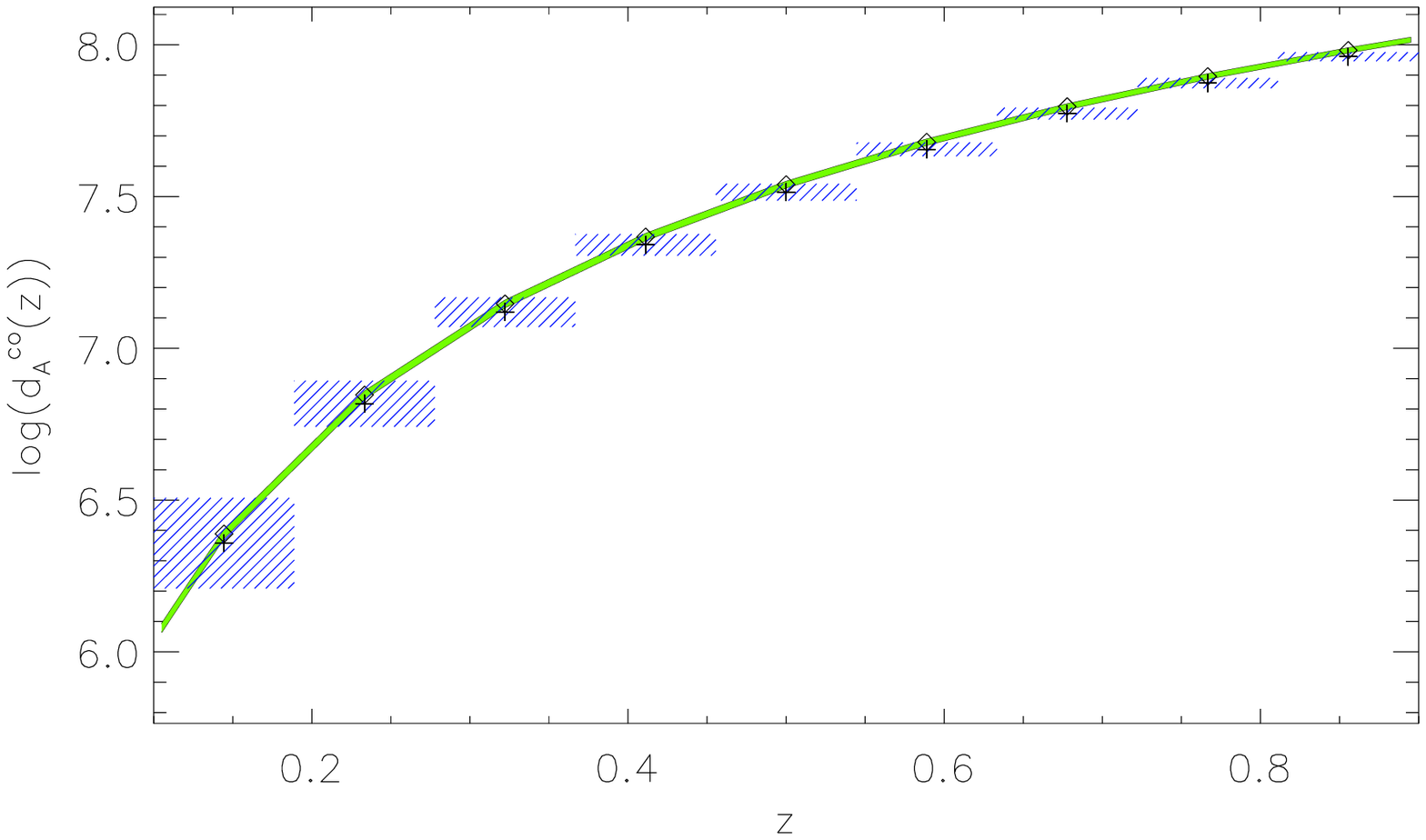}
\hspace{0.5cm}
\includegraphics[width=7cm]{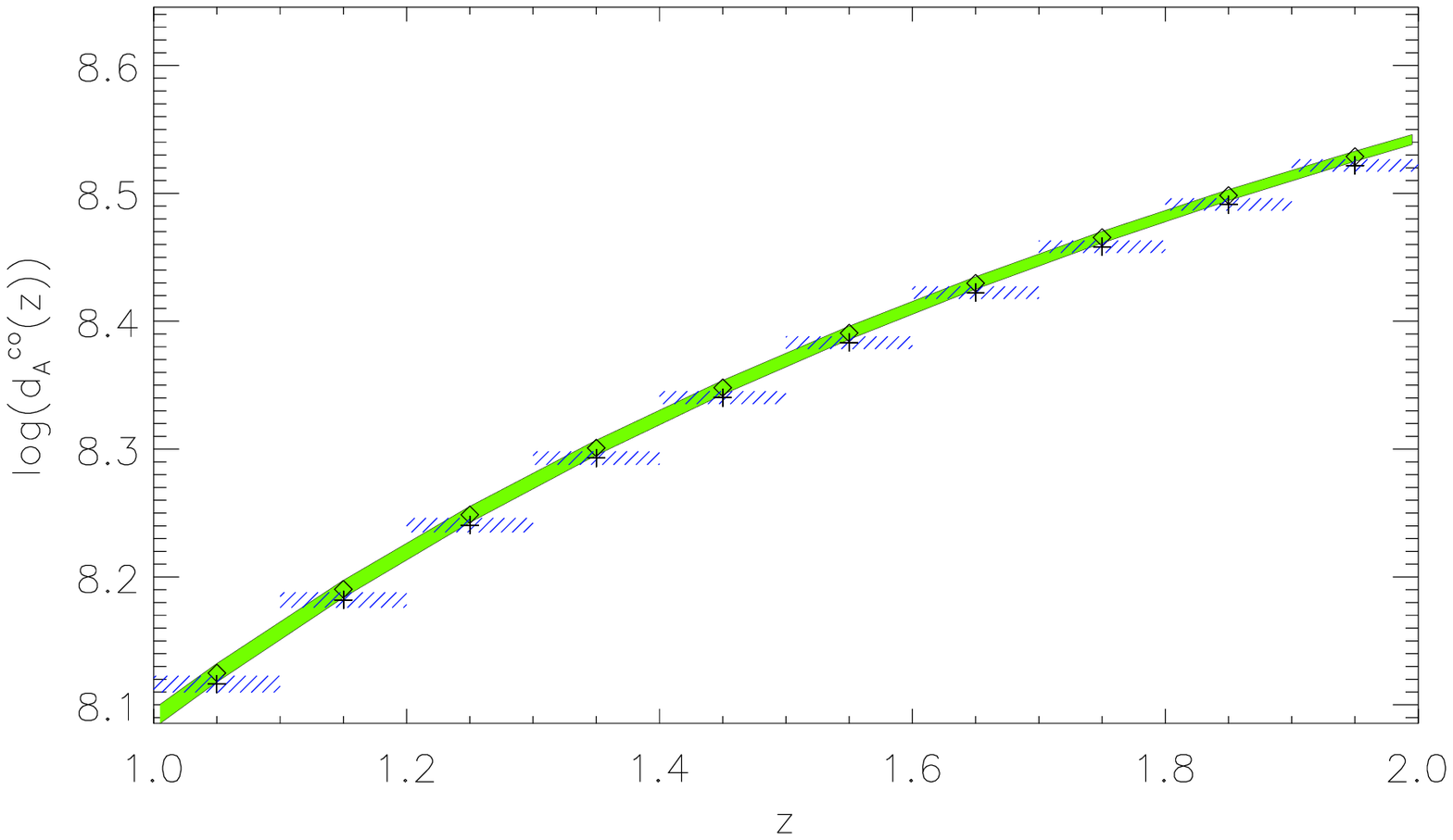}
\includegraphics[width=7cm]{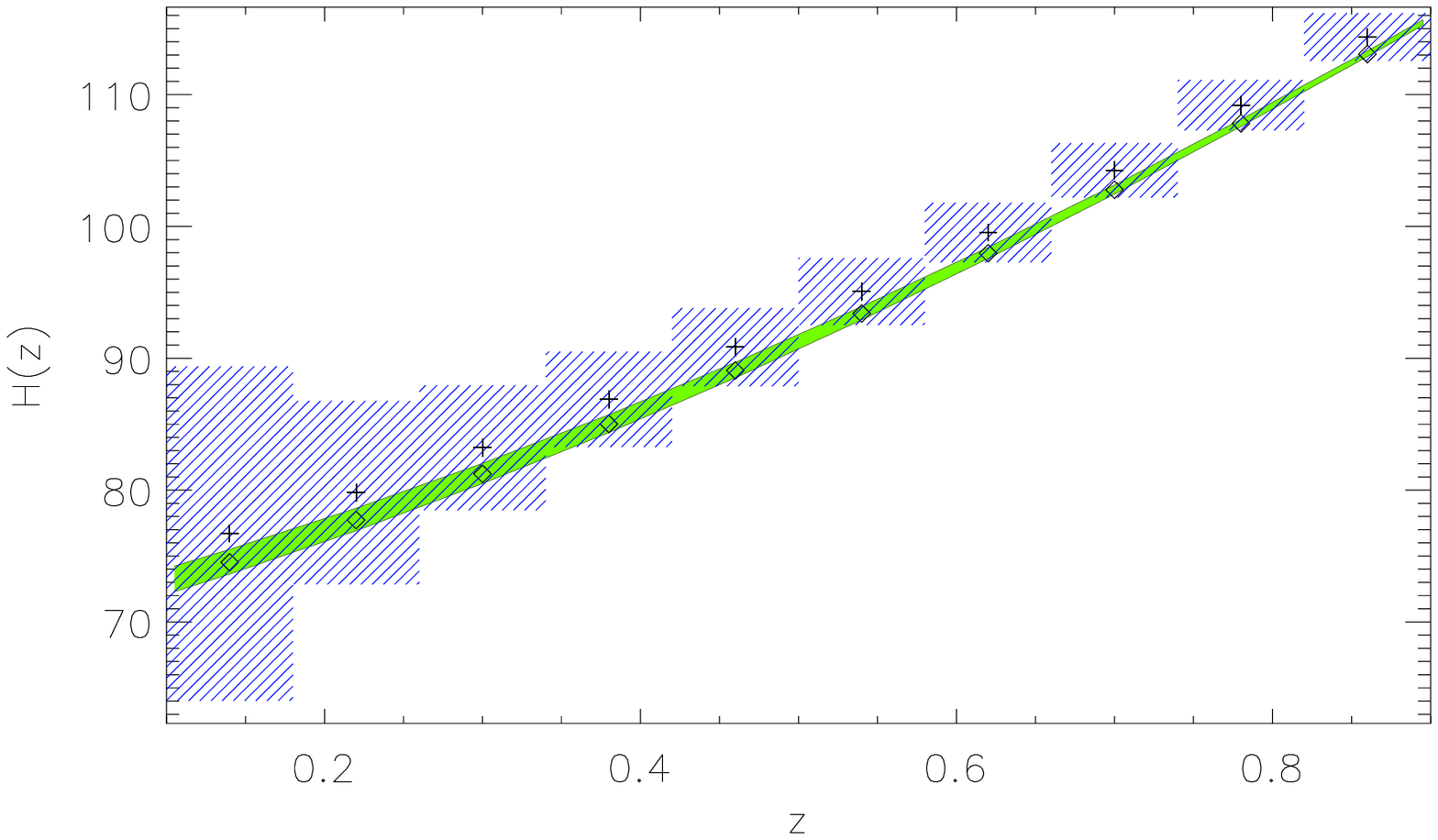}
\hspace{0.5cm}
\includegraphics[width=7cm]{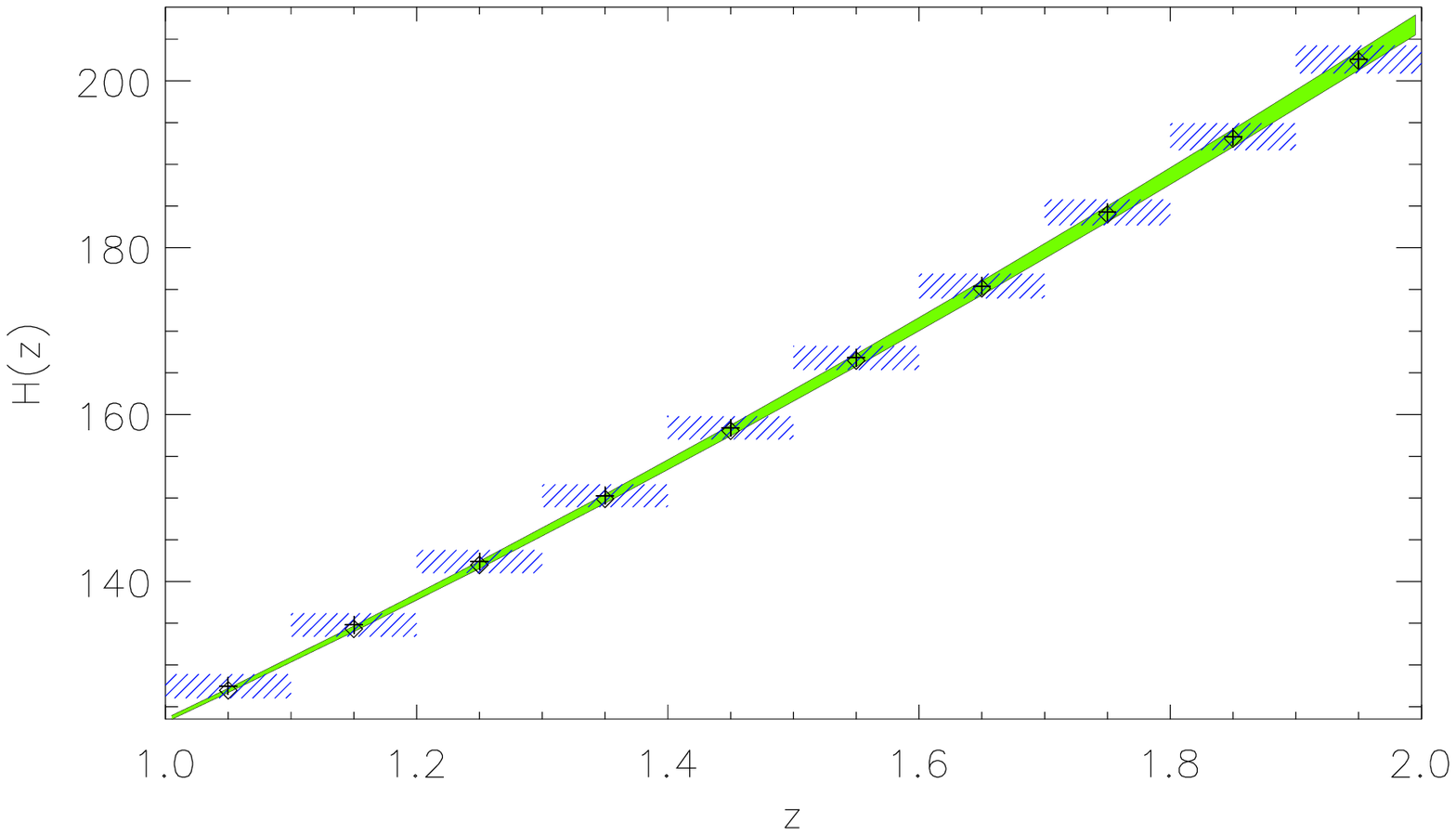}
\includegraphics[width=7cm]{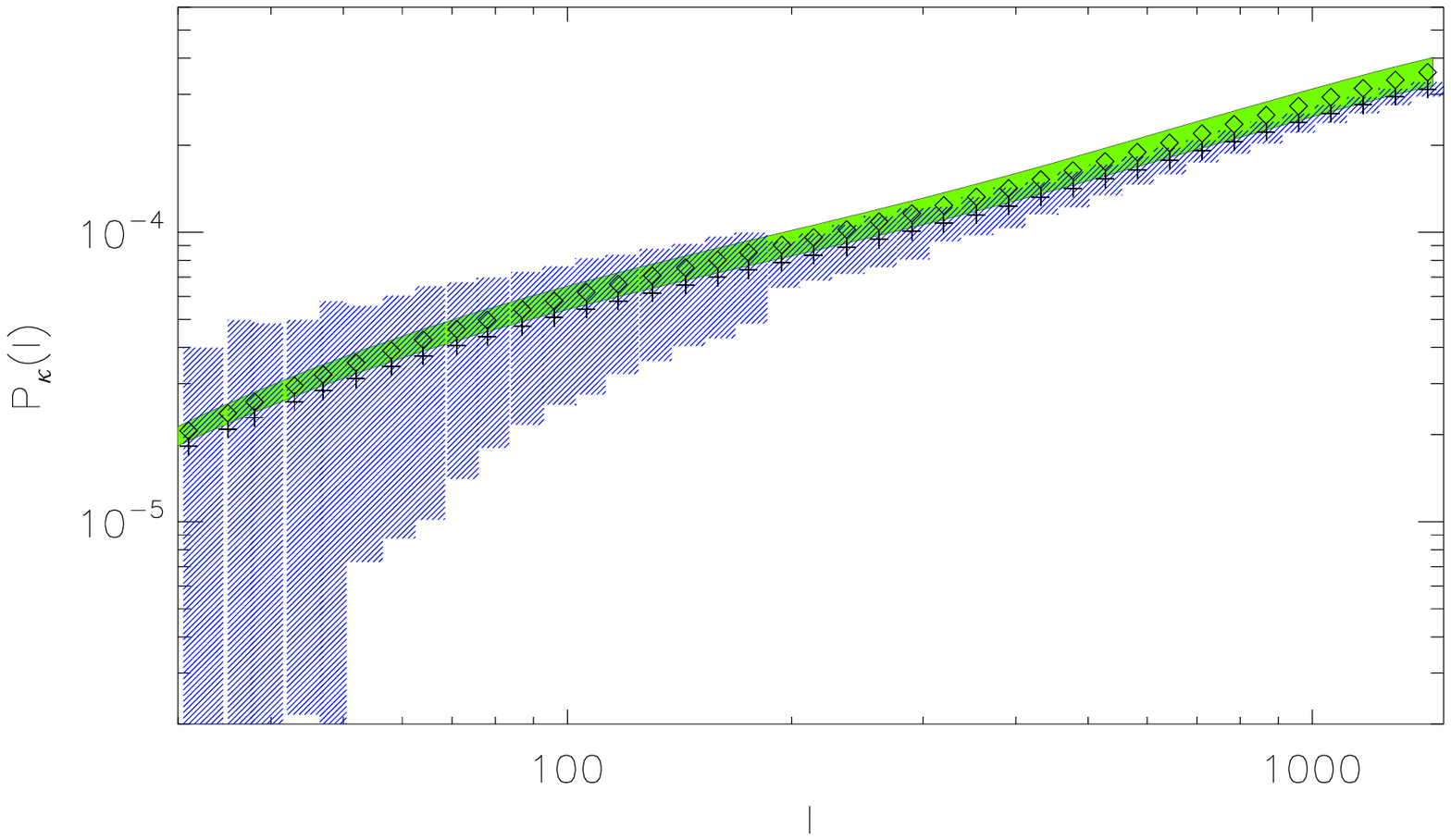}
\hspace{1.5cm}
\includegraphics[width=7cm]{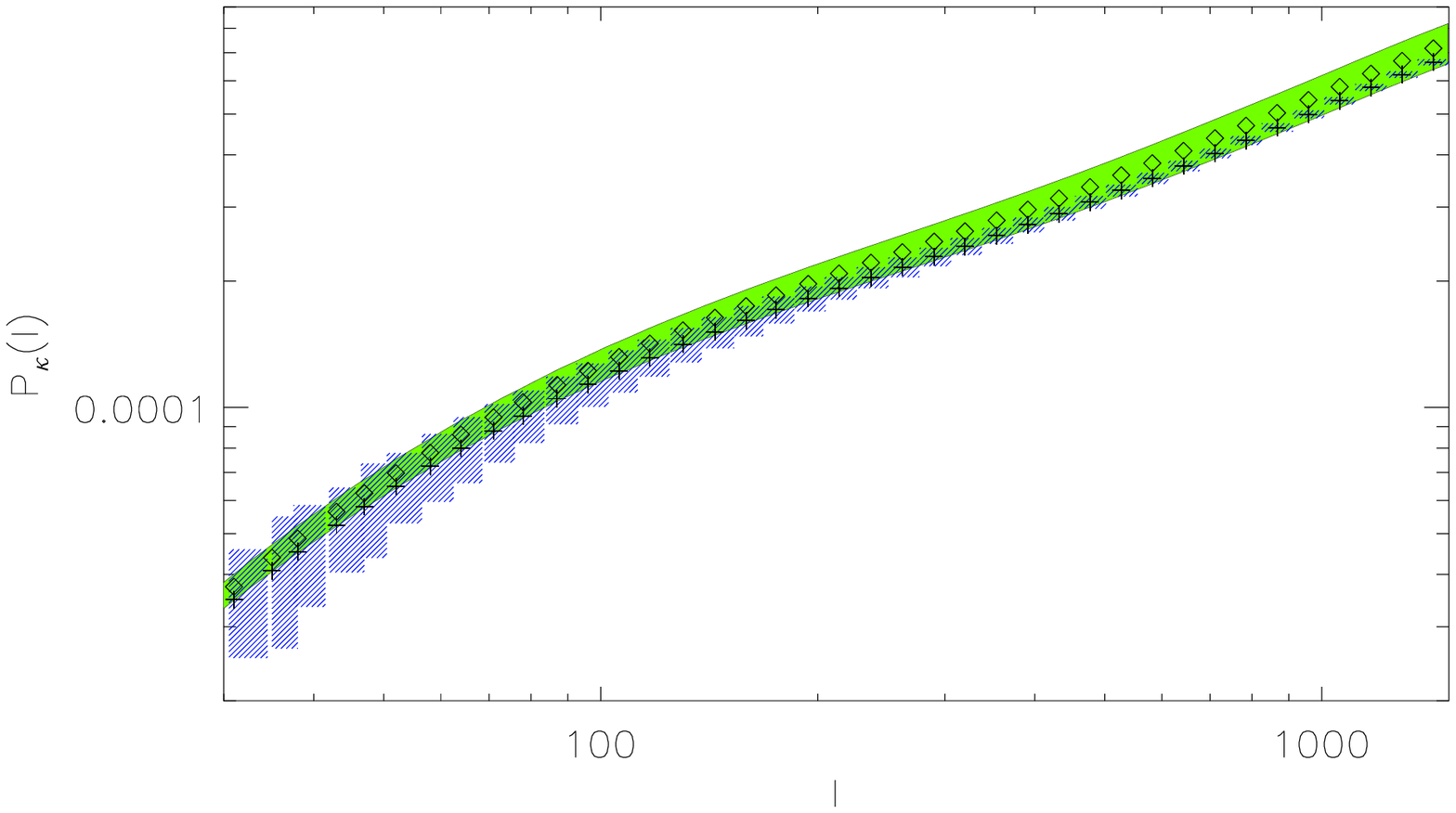}
\caption{Comparison of for ``true'' JBD model (blue) and the extrapolation, assuming GR, of the 1$\sigma$ confidence region from the CMB chain (green). In left panels the JBD model has $\omega_{\rm JBD}$ = 500, in the right panels  $\omega_{\rm JBD}$ = 1000. Top panel: distance modulus, rescaled to an empty Universe, for DETFII (left) and SNAP/LSST (right). Upper middle panel: 
log of the comoving angular distance, for the PAU-BAO (left) and ADEPT (right) surveys. Lower middle panel: Hubble parameter,  for the PAU-BAO (left) and ADEPT (right). In the first case constraints from $H(z)$ are in this case comparable to those from $d_A(z)$, while the two sets of data points lie on top of each other for ADEPT's redshifts, so that $H(z)$ does not carry additional information. Bottom panel: Convergence power spectrum as a function of multipole, for DES (left) and DUNE/LSST (right)}
\label{fig:total}
\end{figure}

For all the late-time observables we have considered, we show two relevant experimental configurations, one for the ``2010'' time line and $\omega_{\rm JBD}$ = 500, and the other for ``2020'', $\omega_{\rm JBD}$ = 1000, in Fig. \ref{fig:total}.

\begin{table}[htbp!]
\begin{center}
\vspace*{0.3cm}
\caption{Specifics of planned and future experiments for Supernov{\ae} type Ia, and statistical significance of forecast detection of $\omega_{\rm JBD}$ = 500, 1000. The minimum redshift is assumed to be 0 for all surveys, and we divide SNe in bins of width 0.1.}
\vspace*{0.3cm}
\begin{tabular}{|l|c|c|c|c|c|}
\hline 
Survey & date av.  & max z &
  SNe tot & $\sigma (\omega_{\rm JBD}$ = 500) & $\sigma (\omega_{\rm JBD}$ = 1000) \\
\hline
DETFII & 2010 & 1.0 & 1200 & 3.1 & 1.4 \\
ONGOING   & 2010 & 0.9 & 850  & 2.9 & 1.4 \\
LSST/SNAP & ``2020'' & 1.7 & 2500 & 7.0 & 3.3 \\
 \hline
\label{tab:sne}
\end{tabular}
\end{center}

\end{table}

\begin{table}[htbp!]
\small
\begin{center}
\vspace*{0.3cm}
\caption{Specifics of planned and future experiments for BAO, and statistical significance of forecast detection of $\omega_{\rm JBD}$ = 500, 1000. Where appropriate, we report in brackets the significance level without using the information from from $H(z)$.}
\vspace*{0.3cm}
\begin{tabular}{|l|c|c|c|c|c|c|}
\hline 
Survey & date av.  & min z & max z &
  ${\rm deg}^2$ & $\sigma (\omega_{\rm JBD}$ = 500)& $\sigma (\omega_{\rm JBD}$ = 1000) \\
\hline
SDSS LRG & 2008 & 0.14 & 0.47 & 7000 & - & - \\
SDSS LRG (BOSS) & 2013 & 0.14 & 0.75 & 7000  & 2.3 (1.8) & - \\
PAU-BAO & 2014 & 0.1 & 0.9 & 10000 & 3.1 (2.6) & 1.2 \\
ADEPT & ``2020'' & 1.0 & 2.0 & 30000 & 7.3 (7.2) & 3.5 (3.4) \\
 \hline
\label{tab:BAO}
\end{tabular}
\end{center}
\normalsize
\end{table}

\begin{table}[htbp!]
\small
\begin{center}
\vspace*{0.3cm}
\caption{Specifics of planned and future experiments for weak lensing, and statistical significance of forecast detection of $\omega_{\rm JBD}$ = 500, 1000.}
\vspace*{0.3cm}
\begin{tabular}{l|c|c|c|c|c|c|c|}
\hline 
Survey & date av.  & med z &
  ${\rm deg}^2$ & {\bf $n/ {\rm arcmin}^2$} & photo-z err & $\sigma (\omega_{\rm JBD}$ = 500)&  $\sigma (\omega_{\rm JBD}$ = 1000) \\
\hline
DES & 2009 & 0.7 & 5000 & 10 & 0.05(1+z) & 3.6 & 2.1 \\
Pan-STARRS & 2012 & 0.7 & 30000 & 5 & 0.06(1+z) & 6.0 & 3.6 \\
LSST/DUNE & ``2020'' & 1.0 & 20000 & 100 & 0.025(1+z) & 6.8 & 3.8 \\
 \hline
\label{tab:WL}
\end{tabular}
\end{center}
\normalsize
\end{table}

\section{Conclusions}
\label{conclu}

We have studied the observational features of Jordan-Brans-Dicke (JBD) theories of gravity, as opposed to General Relativity (GR). Deviations from GR are described by a scalar field non-minimally coupled to  the graviton; when the coupling parameter of the theory, $\omega_{\rm JBD}$, goes to infinity, the GR limit is recovered. We have proposed a method which would allow one to reveal failures of GR without assuming previous knowledge of the true theory of gravity. 

We have first exploited the well-known method of assuming the same expansion history for a JBD and a GR model, and analyzing the associated perturbation growth, which will depend on the underlying theory of gravity \cite{Lue1,Linder:2005in,Koyama:2005kd,Knox_Song_2,Upadhye,Ishak_Spergel,Huterer:2006mv,Kunz:2006ca,Bludman:2007kg}. We showed that this method is not really applicable in scenarios like this, where gravity modifications take place at early times. This suggested the idea of looking for inconsistencies between ``early'' and ``late'' time observables as opposed to expansion history and perturbation growth.  

The method we propose is general and can be used to discriminate between any two theories who give similar predictions at some epoch, and diverge at some other epoch. It only relies on the fact that a correct extrapolation of observations made at early times to late times, or vice-versa, requires the knowledge of the theory of gravity. Even if gravity is GR, it is still possible that the extrapolation gives wrong results if the matter-energy content of the Universe is mistaken, such as in Early \cite{Hebecker:2000zb,ArmendarizPicon:2000dh,Albrecht:1999rm,Wetterich:2003jt} versus Ordinary \cite{Wetterich:1987fm,Ratra:1987rm,Caldwell:1997ii} Quintessence models. \\
We have used  the primary CMB as the early-time observable and    
Supernov{\ae}, Baryon Acoustic Oscillations and  Weak Lensing as the  
late time observables.
Further gain in significance of the results could be obtained considering more observables. The ``early-times'' class is, of course, the more troublesome as for direct observations; in this respect, future surveys of the 21cm hydrogen transition line from the high-redshift IGM \cite{Furlanetto:2006jb} could provide relevant improvements. 

We translated our results into limits on the coupling parameter of the theory, $\omega_{\rm JBD}$. We showed that next-generation experiments will be able to improve substantially the current limit $\omega_{\rm JBD} > 120$ \cite{Acquaviva:2004ti}; detection of values as large as 500 and 1000, respectively, are within reach of the ``2010'' and ``2020'' experiments, for all the observables we considered (SNe, BAOs, Weak Lensing). 

Limits on the JBD parameter can be thought of in terms of an effective parametrization of deviation from GR and in particular can be interpreted as limits on 
the variation of the gravitational constant, G. In fact, we have seen that the evolution of the effective gravitational constant is driven by the evolution of the JBD field, so that these observations constrain the time variation of fundamental constants \cite{Clifton:2004st,Nagata:2003qn}. Constraints of the type $\omega_{\rm JBD} >$ 500, 1000 would correspond to a variation $(|G_{\rm rec} - G_0|)/G_0 <$ 1.38$\%$ and 0.69$\%$ respectively.

We also note that all the observables used in this paper are among those indicated as primary science goals by the recent reports \cite{Peacock:2006kj,DETF}.

To conclude, let us recall that limits on $\omega_{\rm JBD}$ on cosmological scales, as in \cite{Acquaviva:2004ti} and previously in \cite{Damour:1998ae}, were obtained starting from the assumption of JBD as the correct theory of gravity. We also aim to compare constraints obtained with such approach and with the one we used in this paper, on the basis of the same presently available data sets \cite{work_in_prep}.  Such analysis will help to clarify which is the best approach to pursue in the quest for the true theory of gravity.

\section*{Acknowledgments}
We thank Carlo Baccigalupi, Charles Bennett and Alan Heavens for useful comments.
VA is supported in part by NSF grant PIRE-0507768. LV is supported by NASA grant ADP03-0000-009   and ADP04-0000-093.

\label{lastpage}

\bibliographystyle{iopart}
\providecommand{\newblock}{}

\section{Appendix A: Behavior of the matter density contrast in GR and JBD}

\begin{figure}
\includegraphics[width=7cm]{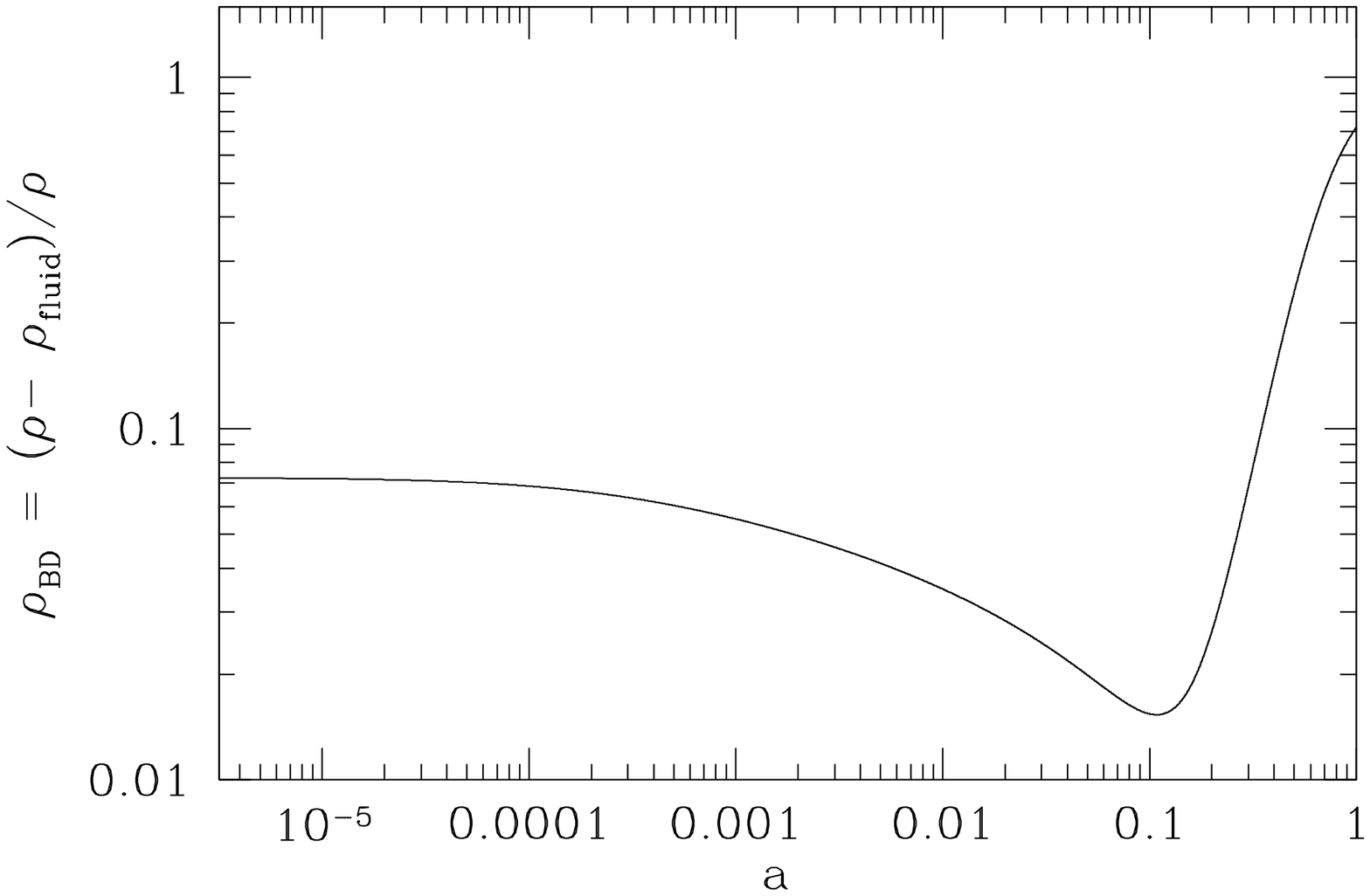}
\hspace{0.5cm}
\includegraphics[width=7cm]{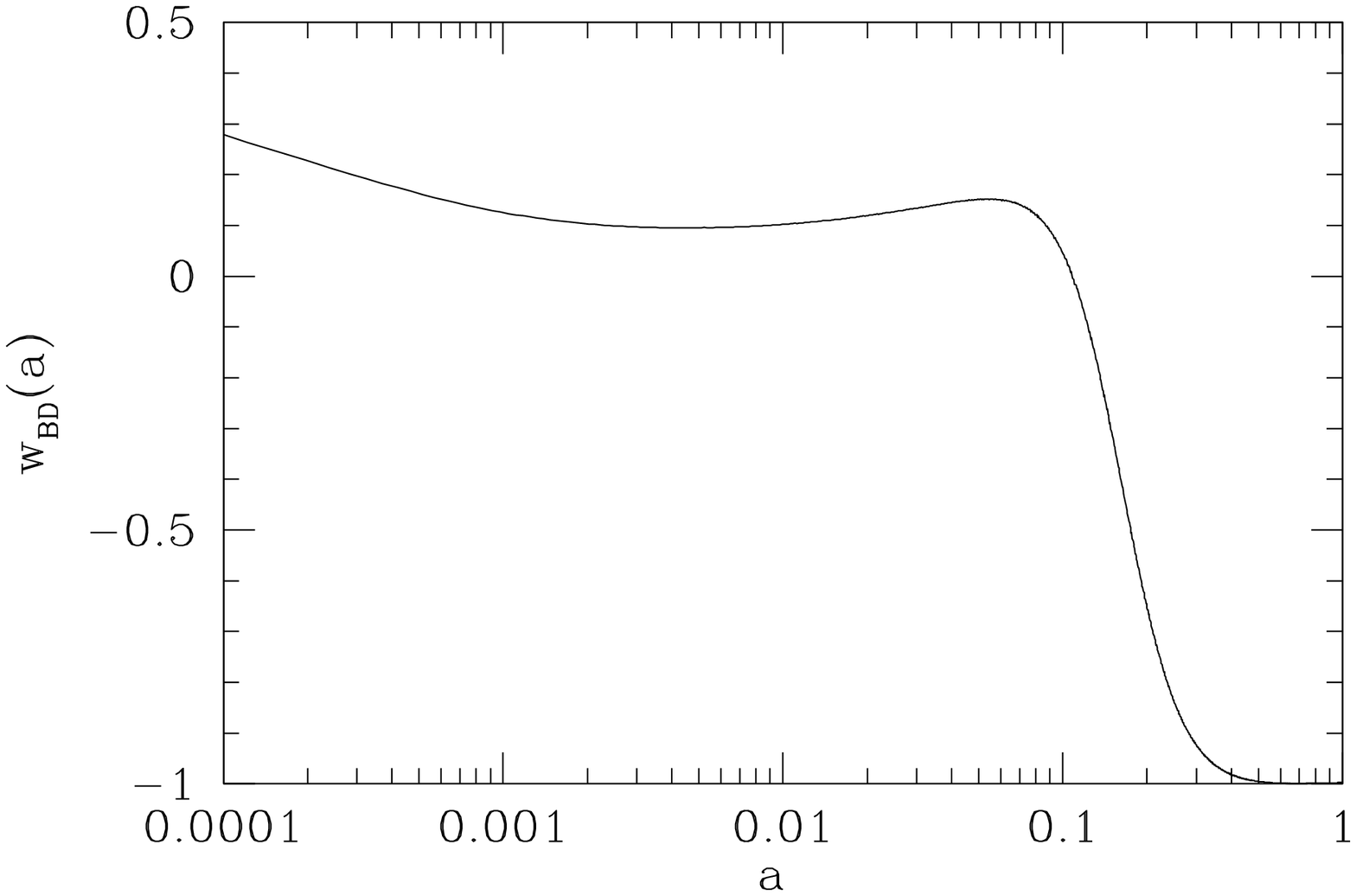}
\caption{Left panel: Relative weight of the hypothetical fluid component in a  minimally coupled model with respect to a JBD model with $\omega_{\rm JBD} = 100$ and coinciding expansion history. Right panel: Equation of state of such fluid component.}
\label{fig:w}
\end{figure}

\begin{figure}
\label{fig:growth}
\includegraphics[width=7cm]{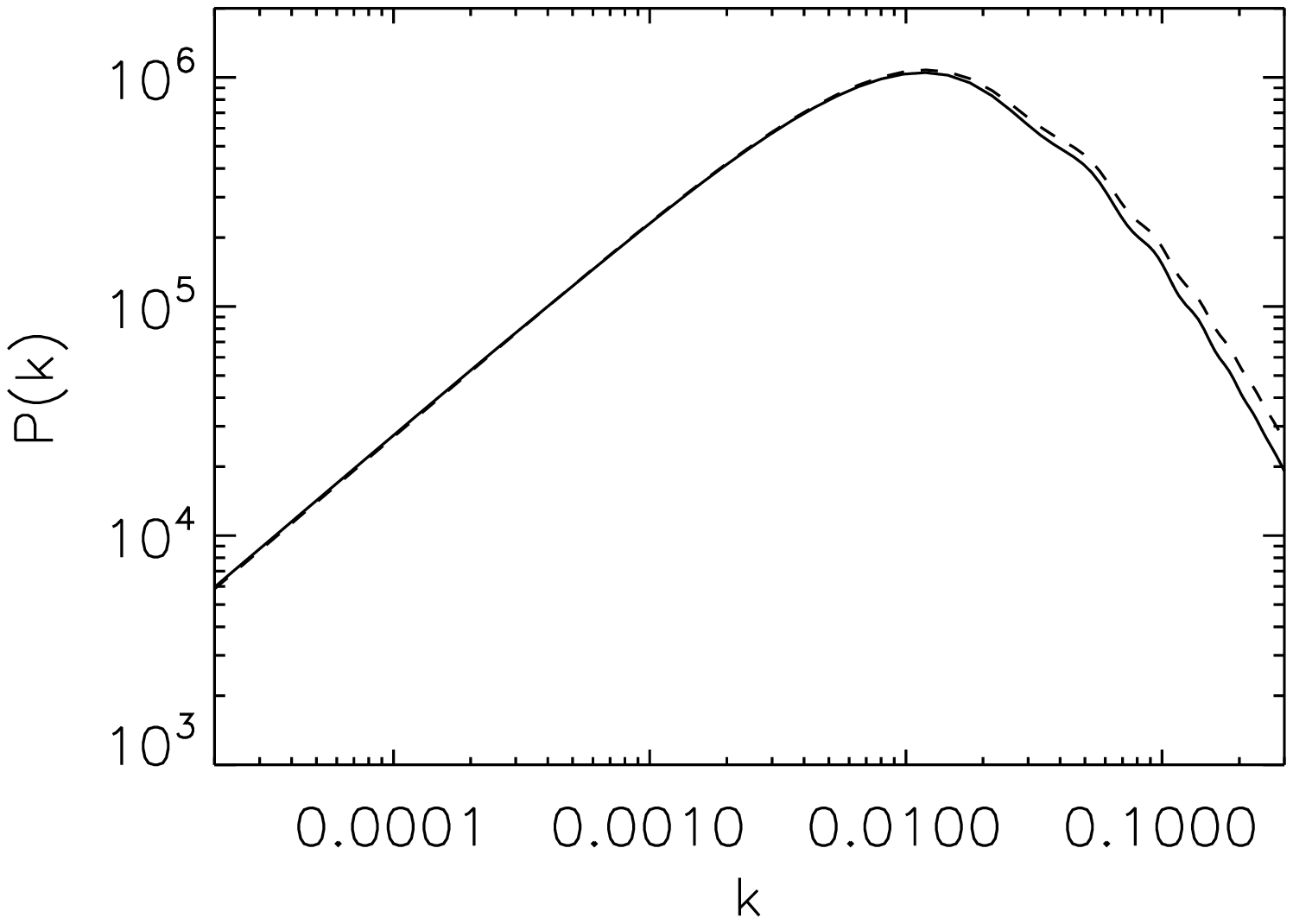}
\hspace{0.5cm}
\includegraphics[width=7cm]{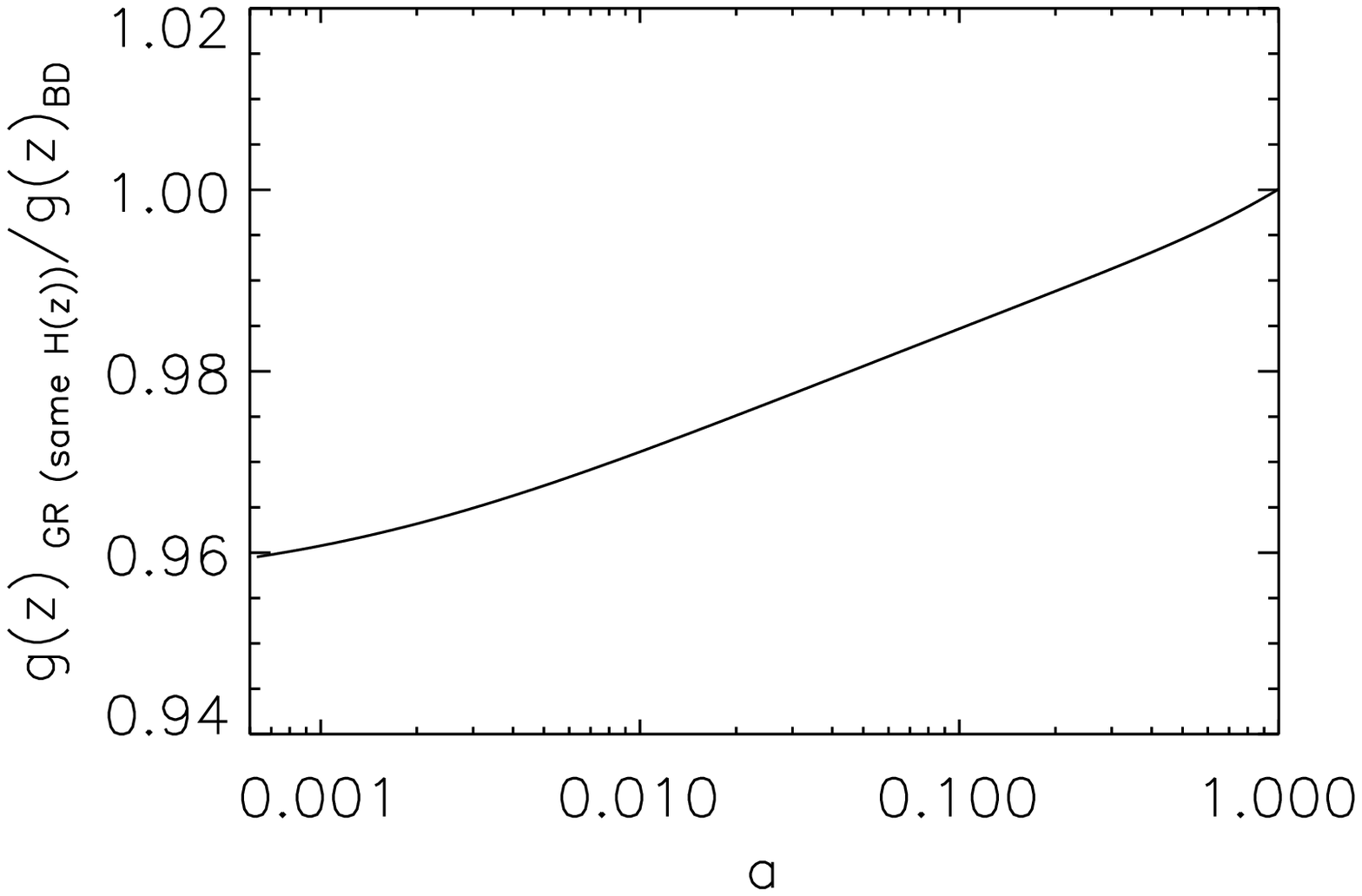}
\caption{Left panel: Linear matter power spectra of the JBD model with $\omega_{\rm JBD} = 100$ (solid line) and of a GR, $\Lambda$CDM model with the same expansion history and cosmological parameters. Right panel: Ratio of the perturbation growth factor g(z) for the same two models.}
\end{figure}

We have seen that our formulation of JBD and GR differ mostly at early time and  that therefore late-time observables, such as the density contrast $\delta_m$,  are not sensitive to $\omega_{JBD}$.  We justify such statement here in more detail. \\
The redshift dependence of $\delta_m(z)$ in GR can be obtained from its evolution equation:
\be
\label{eq:deltam_gr}
\ddot{\delta}_m(z) + 2{H}(z) \dot{\delta}_m(z) = 4\pi G \rho(z) \delta_m(z). 
\ee
$H(z)$ determines $\rho(z)$ exactly through the Friedmann equation, so that the coefficients of the above equation are unambiguously defined.\\
On the other hand, we have learned that  the modifications to GR that we are considering induce a time dependence of the gravitational constant, so that the above equation becomes \cite{Boisseau:2000pr} \footnote {We are indeed neglecting anisotropic stress, which would introduce a non-trivial dependence of the solution on the wavenumber $k$. However, we have numerically checked that the difference of the two gravitational potentials is very small in such models, so that the error in this approximation is $< 10^{-4}$.}
\be
\label{eq:deltam_bd}
\ddot{\delta}_m(z) + 2{H}(z) \dot{\delta}_m(z) = 4\pi G_{\rm eff}(z) \rho(z) \delta_m(z). 
\ee
The solution for $\delta_m(z)$ in this case, even assuming the same $H(z)$, will be different from GR for two reasons: first, the mapping between $H(z)$ and $\rho(z)$ will change, as seen in the previous section, and second, the source term of such equation will also change, in response to the time variation of $G_{\rm eff}(z)$. \\
 However, how well can we expect this method to do for JBD models? The function $G_{\rm eff}$(z) in this case is $1/(8 \pi F)$: this implies, given the evolution of $1/F$ from Fig. \ref{fig:trajectory}, that the time dependence of $H(z), \rho(z)$ and $G_{\rm eff}(z)$ will differ significantly from the ordinary gravity case only at high redshift. At redshift, say, $z < 1$, the coefficients of the differential equations (\ref{eq:deltam_gr}) and  (\ref{eq:deltam_bd}) will be similar, and so will be the solution $\delta_m$.  \\
 To illustrate all this,  we have developed a method to reproduce exactly the expansion history of a given JBD model, which we will call $H(z)_{\rm ref}$, in an Ordinary Gravity case.  This cannot be done by simply changing the cosmological parameters: in a flat $\Lambda$ CDM model $H(z)$ only depends on $h_0$ and $\Omega_m$ (and $\Omega_r$  for $z>>100$),
\be
\label{HzGR}
H_{\rm GR}(z)=H_0  \frac{1}{\sqrt{\Omega_r (1+z)^4 + \Omega_m (1+z)^3 + \Omega_{\Lambda}}}\,,
\ee
But in the JBD case there is no such simple functional form of the Hubble factor;
the complicated evolution of ${H_{\rm ref}}(z)$, given by Eq. \ref{eq:h}, cannot be described as a sum of the three power-laws in redshift which appear in the equation above. \\
 We consider a GR, three-fluid model with  ordinary matter and radiation components plus a perfect fluid with  equation of state $w_{\rm JBD}$(z). We will refer to such component as the ``Jordan-Brans-Dicke fluid'', and other than the value of $w_{\rm JBD}$ we will impose the same cosmological parameters as the "true" JBD model. This way, we expect to isolate the effect of the JBD field at early times and to recover the standard Cosmological Constant term at late times.
For this model, the expression for $H(z)$  is identical to Eq. (\ref{HzGR}) but with the substitution $\Omega_{\Lambda} \longrightarrow \Omega_{\rm JBD}(z)$
where
\be
\label{eq:ombd}
\Omega_{\rm JBD}(z)= \Omega_{\rm JBD}(z=0) \exp\left[-3\int^0_z (1+w_{\rm JBD}(z))dz\right];
\ee
$\Omega_{\rm JBD}(z=0)$ is the present value of the JBD field density, and can be fixed as $1 - \Omega_m$ requiring geometrical flatness.\\
With a generic $w_{\rm JBD}$(z), any function $H(z)$ can be exactly reproduced in this form. 
The relative weight of the ``JBD fluid'' density measures the contribution of the JBD field to the total energy density, and therefore to the expansion history. It can be evaluated as $(\rho_{\rm JBD} - \rho_{\rm tot})/\rho_{\rm tot}$, and is shown in the left panel of Fig. \ref{fig:w} for the expansion history of a model with $\omega_{\rm JBD} = 100$. On the right panel we show the equation of state of the fluid, $w_{\rm JBD}$, which, is helpful for the interpretation of physical effects of the JBD field.
At very early times the relative contribution of such component is five to ten percent for this value of $\omega_{\rm JBD}$, and its equation of state is positive and in the range between 0.2 and 0.3. Such fluid may be interpreted as a mixture of matter and radiation. \\
At the CMB redshift, $z \simeq 1080$, the equation of state of the JBD fluid is close to zero, mimicking an additional matter component, which is indeed expected from the behavior of the multiplicative term $1/F$ in the density equation. We have seen that at early times the gravitational coupling in the JBD model is stronger than in GR; in order to reproduce this feature in the Ordinary Gravity case, one would need to enhance the total matter density. \\ 
At late times the fluid behaves, as anticipated, as a Cosmological Constant component, and its equation of state at  $z < 1$ is practically indistinguishable from $-1$ . Therefore, at late times we cannot expect such fluid to contribute to the evolution of $\delta_m$. 
To illustrate this, in Fig. \ref{fig:growth} we plot the power spectrum of matter, evaluated at the present time, for the two models, and the ratio of their perturbation growth factors. Although for wavenumbers $k \geq 0.1$ Mpc$^{-1}$ there is a relevant difference between the two power spectra, it has to be attributed to the difference in the transfer function. The field behaves as an additional matter component, therefore the horizon size at matter-radiation equality changes, and so does the transfer function on smaller scales. The redshift evolution of the growth factor $g(z)$ in the two cases only shows differences of $\leq 4\%$, and, as expected, mainly at redshift higher than those probed by large-scale structure or lensing surveys. Furthermore, let us stress that the value of $\omega_{\rm JBD}$ we are considering for illustrative purposes is fairly large and already ruled out by cosmological probes \cite{Acquaviva:2004ti}.  It is indeed true that while techniques used so far to trace $\delta_m$, such as  galaxy surveys, weak lensing, cluster counts  have only been used at z $\leq$ 1, different methods may be used to probe much larger redshifts. The proposed technique of tracing the neutral Hydrogen using its 21 cm transition is challenging (e.g. \cite{Furlanetto:2006jb}), but very promising and may provide accurate maps of the high redshift large scale dark matter distribution.  However, for the time being, the detection of deviations from GR for a JBD model through inconsistencies in the expansion history and perturbations growth are limited by the poor measurements of $\delta_m$ at early times.

\section{Appendix B: Code comparison}
\begin{figure}
\includegraphics[width=7cm]{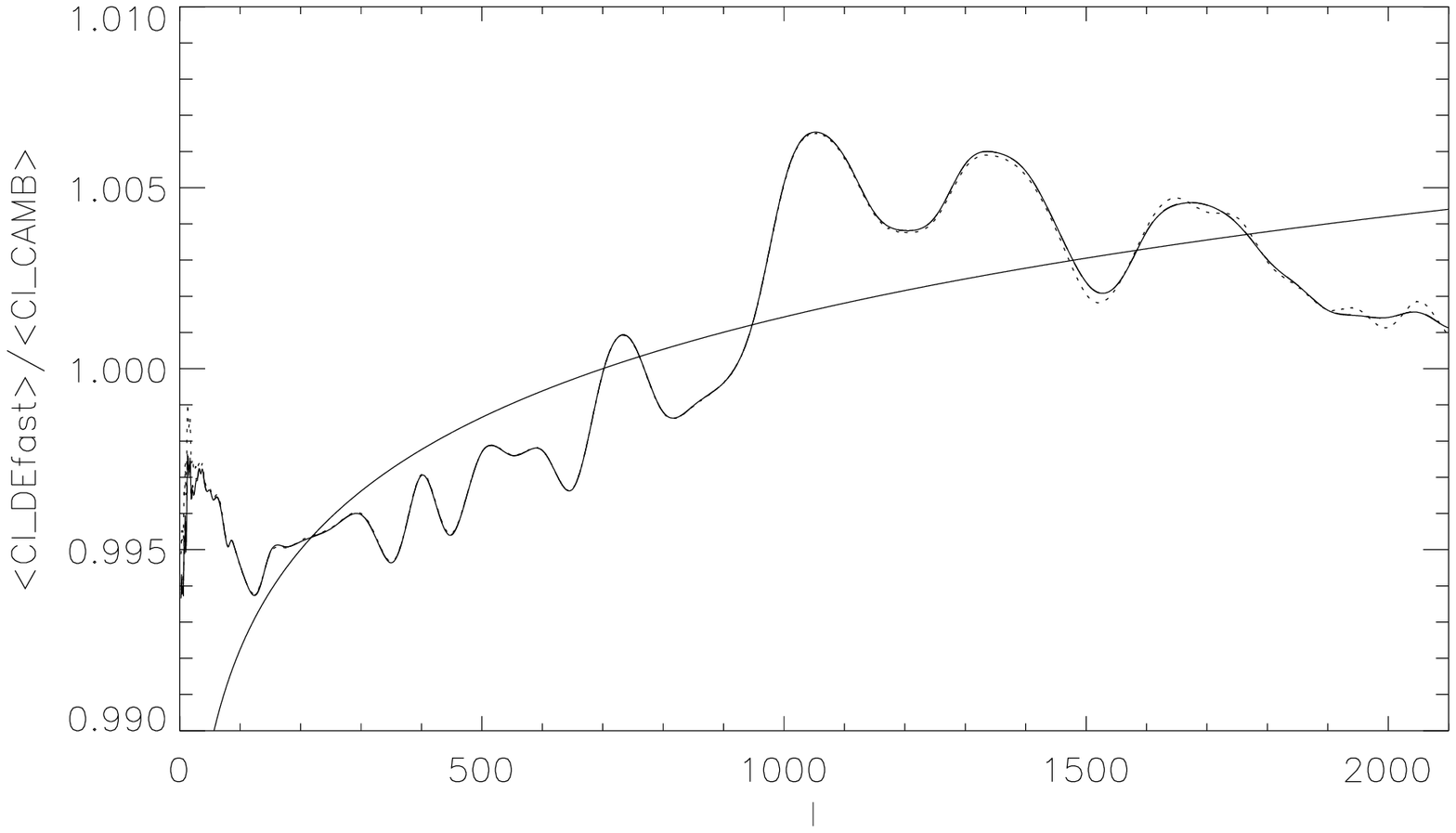}
\hspace{0.5cm}
\includegraphics[width=7cm]{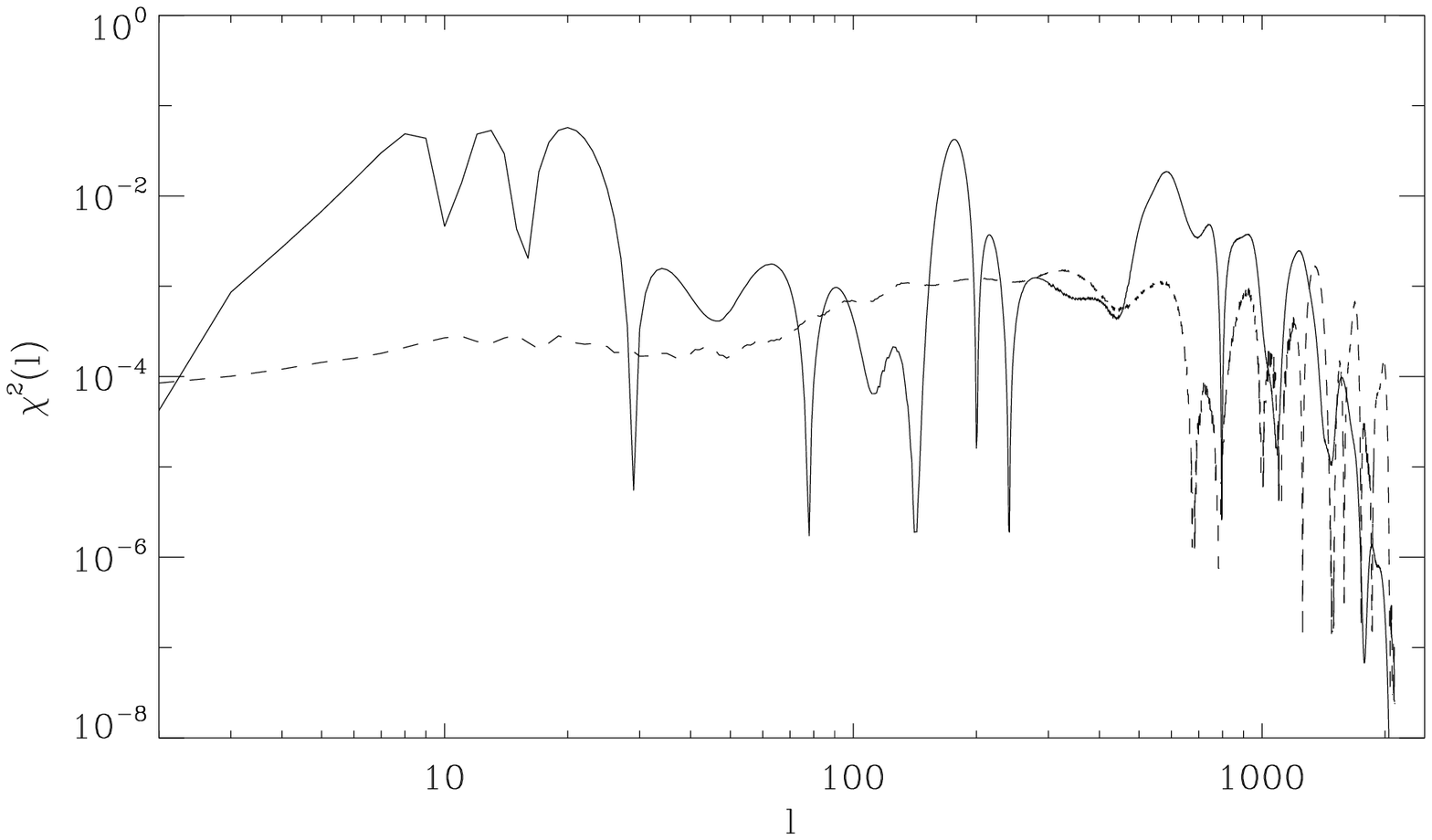}
\caption{Left panel: Ratio between the temperature power spectra as a function of multipole, for 10 (dotted line), 100 (dashed line), and 520 (solid line) sample models. \quad Right panel: distribution of the $\chi^2$ due to the disagreement between DEfast and CAMB, for the temperature (dashed line) and EE polarization (solid line) spectra, as a function of multipole.}
\label{fig:camb-defast}
\end{figure}

The DEfast code, used to produce the reference JBD spectra, and the CAMB code used by the COSMOMC program are known to be in agreement within the $1\%$ level \cite{tesi_ilaria}. However, the shift in the cosmological parameters we are looking for are of comparable size. In order to properly calibrate outputs, we run several chains, for different values of the cosmological parameters, for GR $\Lambda$CDM models, using input spectra from DEfast. Only if all the values of the cosmological parameters are recovered correctly one can be sure that the observed shift is due to the JBD nature of gravity rather than to a numerical effect. We report in the second column of Table \ref{table:shift} the result of one such chains, for the same reference model quoted in the main text. It can be seen that, for all the parameters, the values resulting from the fit lie within the $1-\sigma$ vicinity of the true ones. However, we noticed a slight positive shift in the reconstructed value of $n_s$, of the order of $0.4\%$, which appeared to be of systematic nature. We thus compared the output of the DEfast code and the CAMB code for a sample of 520 models of the chain mentioned above, indeed finding some residual numerical difference. In particular, fixing the primordial amplitude normalization scale at $k  \simeq 0.05$ $\rm Mpc^{-1}$, corresponding to $l_0 \simeq 700$, the DEfast code predicts a slightly lower power in temperature, with respect to CAMB, at lower multipoles, and slightly more power at higher multipoles. We however discovered that such difference is largely independent of the cosmological model: we averaged the ratio of the power spectra given by CAMB and DEfast over N = 10, 100 and 520 models and found extremely similar curves, as plotted in Fig. \ref{fig:camb-defast}. Let us notice that the shape of such difference is very close to what would be caused by a shift in the value of the spectral index of the order of the one we observe; we plot the corresponding effect on the power spectrum:
\be
\frac{C^{TT}_l(n_s)}{C^{TT}_l(n_s + \delta n_s)} = \left(\frac{l}{l_0}\right)^{- \delta n_s},
\ee
 to be compared with our empirical correction curve, in the same Figure (smooth solid line). \\ 
We concluded that the ratio of the power spectra from the two codes, averaged over a large number of models, can be used as an effective correction to the DEfast temperature power spectrum. Results of a chain run with the same cosmological parameters and the corrected $C_l^{TT}$ are shown in the third column of Table \ref{table:shift}: we see that the shift in the fitted value of $n_s$ is now negligible. This correction was used throughout all our numerical analysis. \\
However, even after this correction, some small differences in the output from DEfast and CAMB remained. Even if not harmful for our procedure of reconstructing cosmological parameters, they are responsible for the fact that the value $\chi^2$ for the reconstructed model does not approach zero as we recover GR. In fact, the fit to a GR $\Lambda$CDM model whose input power spectra are generated with DEfast has a value of $\chi^2$ of 9.06; this is due to the fact that COSMOMC uses CAMB in order to generate the CMB spectra. The same fitting procedure, if done using input spectra generated by CAMB, would yield a much smaller $\chi^2$ of 0.26. We checked the distribution in the multipole space of the $\chi^2$, finding that it is in large part due to slight differences in the EE polarization spectrum at low l. The total value of the $\chi^2$ for this spectrum alone would be 5.15, while the temperature power spectrum would only have 0.91 (and their cross-correlation, of course, is responsible for the missing 3.02). We show the distribution of the $\chi^2$,  for the TT and EE spectra in the right panel of Fig. \ref{fig:camb-defast}. 

\begin{table}[htbp!]
\begin{center}
\vspace*{0.3cm}
\caption{Best fit and $1 \sigma$ confidence levels of cosmological parameters, before and after correction.}
\vspace*{0.3cm}
\begin{tabular}{l|c|c|c}
\hline 
&  ``true'' & uncorrected & corrected \\
\hline
 $ { \omega_b }$ & $0.022$ &  $0.022^{\, 0.0222}_{\,0.0218}$ & $0.022^{\,0.0222}_{\,0.0217}$ \\
 ${ \omega_{\rm CDM} }$ &  $0.1232$ & $0.1235^{\, 0.1263}_{\, 0.1212}$ & $0.1236^{\,0.1260}_{\,0.1210} $\\
 ${n_s} $ & $0.95$ & $0.9540^{\,0.9602}_{\,0.9482}$ & $0.9511^{\,0.9571}_{\,0.9448}$  \\
 ${ \tau}$ & $0.09 $ & $0.0921^{\,0.0997}_{\,0.0829}$ & $0.0915^{\, 0.0993}_{0.0829}$ \\
 ${\cal A}$ & $3.1355 $ & $3.1400^{\,3.1562}_{\, 3.1221} $ & $3.1411^{\, 3.1572}_{\, 3.1233}$  \\
 ${ h_0 }$ & $0.72 $ & $ 0.7194^{0.7296}_{0.7079} $ & $0.7191^{\,0.7305}_{\, 0.7085}$  \\
 ${ \Omega_m }$ & $0.28 $ & $0.2812^{\, 0.2955}_{\,0.2695}  $ & $0.2815^{\,0.2946}_{\, 0.2683}$  \\
 \hline

\end{tabular}
\label{table:shift}
\end{center}
\end{table}

\end{document}